\documentclass[draft]{agujournal2019}
\usepackage{url} 
\usepackage{lineno}
\usepackage[inline]{trackchanges} 
\usepackage{soul}

\newcommand{\CO}{CO$_2$ }
\newcommand{\HHO}{H$_2$O }

\newcommand{\Ls}{L$_s$}

\newcommand{\sigO}{1-$\sigma$ }
\newcommand{\tiu}{J~m$^{-2}$~K$^{-1}$~s$^{1/2}$}
\draftfalse

\journalname{JGR Planets}

\begin{document}

%
%


\title{Modelling Slope Microclimates  in the Mars Planetary Climate Model}

%
%




\authors{L.Lange\affil{1}, F.Forget\affil{1}, E.Dupont\affil{1}, R.Vandemeulebrouck\affil{1}, A.Spiga\affil{1},   E.Millour\affil{1}, M.Vincendon\affil{2}, A.Bierjon\affil{1} }


\affiliation{1}{Laboratoire de Météorologie Dynamique, Institut Pierre-Simon Laplace (LMD/IPSL), Sorbonne Université, Centre National de la Recherche Scientifique (CNRS), École Polytechnique, École Normale Supérieure (ENS), Paris, France}
\affiliation{2}{Institut d’Astrophysique Spatiale, Université Paris-Saclay, CNRS, Orsay, France}



\correspondingauthor{Lucas Lange}{lucas.lange@lmd.ipsl.fr}



\begin{keypoints}
\item We develop a sub-grid slope parameterization to simulate slope microclimates in the Mars Planetary Climate Model;
\item We test and validate this parameterization against observations; and reappraise frost thicknesses on Martian slopes; 
\item This novel model opens the way to new studies on surface-atmosphere interactions for the present and past climates of Mars.
\end{keypoints}

%
%

%
%
\begin{abstract}

A large number of surface features (e.g., frost, gullies, slope streaks, recurring slope lineae) are observed on Martian slopes. Their activity is often associated with the specific microclimates on these slopes, which have been mostly studied with one-dimensional radiative balance models to date. We develop here a parameterization to simulate these microclimates in 3D Global Climate Models. We first demonstrate that any Martian slope can be thermally represented by a poleward or equatorward slope, i.e., the daily average, minimum, and maximum surface temperatures depend on the North-South component of the slope. Based on this observation, we implement here a subgrid-scale parameterization to represent slope microclimates (radiative fluxes, volatile condensation, ignoring slope winds for now) in the Mars Planetary Climate Model and  validate it through comparisons with surface temperature  measurements and frost detections on sloped terrains. With this new model, we show that slope microclimates do not have a significant impact on the seasonal CO$_2$ and H$_2$O cycles. Furthermore,  short-scale slopes do not significantly impact the thermal  state of the atmosphere. 91\% of the active gullies are found where our model predicts CO$_2$ frost, suggesting that their activity is related to processes involving CO$_2$  ice.  However, the low thicknesses ($\leq$~tens of cm) predicted at mid-latitudes there rule out mechanisms involving large  amounts ($\sim$ meters) of ice. This model opens the way to new studies on surface-atmosphere interactions in present and past climates. 

\end{abstract}

\section*{Plain Language Summary}
 Martian slopes present traces of present and past activities  (e.g., dust avalanches, moraines created by past glaciers). Such activities are linked to the microclimate of these slopes: a slope oriented toward the pole will be more in the shadow and therefore colder than a slope oriented toward the equator. For now, mostly 1D models have been used to determine the microclimates present on these slopes, neglecting potential 3D atmospheric  effects. Here we propose a parameterization allowing us to model such microclimates in 3D Global Climate Models. We have implemented and validated this parameterization in the Mars Planetary Climate Model by comparing our model with observations and measurements of these microclimates on the slopes. Our model shows that these slope microclimates do not significantly influence the global climate of Mars. We also provide a new  CO$_2$ ice thickness  map which constrains the possible mechanisms that form gullies on Mars.
%
%

%


%
%
%
%

\section{Introduction}

Martian slopes are associated with a wide range of present-day surface processes \cite{Diniega2021, Dundas2021}.  Gullies,  surface features  composed of an alcove, a channel, and a depositional apron  similar to small terrestrial alluvial fans, are found on  slopes at mid and high latitudes \cite{Malin2000}. At low to mid-latitudes, dark wedge-shaped features on high-albedo steep slopes are associated with downslope mass movements of the first micrometers of the surface, and are known as Slope Streaks  \cite{Sullivan2001}. Dark features called Recurring Slope Lineae (RSL)  are also found on steep slopes that have a low albedo  \cite{McEwen2011, McEwen2014}. Other mass movements reported on Martian slopes include rockfalls and small displacements of high-latitude rocks \cite{Dundas2019}, polar avalanches \cite{Russell2008}, and dune-slope activity \cite{Diniega2019, Diniega2021}.

Although the origin of these features is still debated today, most of the  mechanisms proposed to explain their activity are associated with the microclimates (surface/subsurface temperatures, condensation or sublimation of volatiles, near-surface winds) found on the slopes \cite<e.g., >{Diniega2021,Dundas2021}. Because of the low thickness of the Martian atmosphere today, surface temperatures and near-surface environment are mostly controlled by radiative fluxes at the surface \cite{Savi:08}.  These fluxes are significantly modified for a slope compared to a flat surface.   For instance, pole-facing slopes receive less solar insolation and are thus significantly colder than equator-facing slopes. Therefore, ice should preferentially accumulate on these slopes, as confirmed by current frost observations \cite{Schorghofer2006, Carrozzo2009, Vincendon2010water, Vincendon2015, Dundas2019gullies, Lange2022a}. Consequently, sublimation-driven processes have been proposed to explain some pole-facing slope features that are consistent with the presence of \CO or \HHO ice \cite<see>[for a complete review]{Diniega2021}. Similarly, for slopes reaching the warmest temperatures, i.e., equator-facing slopes, thermally-stress-induced flows have been proposed to explain equator-facing slope features \cite<e.g.>[for RSL]{Schmidt2017, Tesson2020}.

Slopes also present traces of past climates of Mars.  For instance, \citeA{Kreslavsky2011} reported moraines interpreted as remnants of \CO glaciers  on steep pole-facing slopes at high latitudes. These glaciers could correspond to periods of low obliquity when pole-facing slopes are cold traps that favor the condensation of the atmosphere \cite{Kreslavsky2005}. Conversely, at higher obliquity, around the summer solstice,  surface and soil temperature of poleward-facing slopes could possibly reach the melting point of water, melting near-surface ice, possibly creating Martian debris flows features observed at mid-latitudes \cite{Costard2002}.

The consequences of slope microclimates are not limited to the surface as they also influence the thermal environment of the Martian subsurface and its contents.  The Gamma Ray Spectrometer \cite<GRS,>[]{Boynton2004} onboard Mars Odyssey has revealed the presence of subsurface water ice covered by a dry layer of regolith above $\pm$ 55\textdegree~latitude, and locally down to 45\textdegree~\cite{Boynton2002}, which was predicted/globally reproduced by numerical models \cite{Leighton1966, Mellon1995, Mellon1997, Mellon2004, Schorghofer2005}. As pole-facing slopes are colder than surrounding flat surfaces, \citeA{Schorghofer2005} suggested that  subsurface water ice (either pore-filling or excess ice)  could be stable at equatorward latitudes beneath these pole-facing slopes. Local high-resolution subsurface ice mapping indeed showed that the depth of the ice table was correlated to topographic heterogeneity \cite{Bandfield2007,Piqueux2019}. \citeA{Aharonson2006} modeled the theoretical stability of subsurface water ice,  considering the topographic heterogeneity at high resolution, by using a 1D thermal model that accounts for sloped terrains. They found that ice could be stable at latitude as close as $\pm 25$\textdegree~compared to $\pm 45$\textdegree~for former models that do not consider slopes and their particular thermal behavior. \citeA{Vincendon2010} also suggested that the absence of \CO frost at low latitudes on steep south-facing slopes in the Southern hemisphere during winter could be explained by the presence of this subsurface water ice beneath these slopes. Such results have profound implications, especially as shallow water ice can be an exploitable resource for future crewed exploration  \cite{Morgan2021}.

Therefore slope microclimates are necessary to understand current surface and subsurface processes, as well as past climates of Mars. These processes are generally found on steep ($\geq$ 20$^\circ$) slopes that have a typical length of kilometers/tens of kilometers \cite{Kreslavsky1999}. Such short-scale topographies and associated microclimates cannot be represented in a 3D Global Circulation Model (GCM) like the Mars Planetary Climate Model \cite<Mars PCM, formerly known as the LMD-Mars GCM,>{Forget1999} which has a typical resolution of  300 km in longitude and 220 km in latitude at the equator. Mesoscale models with resolutions of tens of kilometers, either on a limited area or on the whole planetary sphere, cannot be used for long-term surface-atmosphere interactions (e.g., glacier evolution following obliquity variations over thousands of years) given prohibitive computational cost.

1D radiative equilibrium models are often used  to study the environment on a slope   \cite<e.g., >{Costard2002,Vincendon2010,Vincendon2010water}.  However, in most of these models, feedback on the large-scale properties of the meteorology is neglected while some surface processes are dependent on global circulations. For instance, surface water ice formation depends on the amount of water vapor and ice precipitation which need to be resolved by a GCM. For such studies,  1D models have been paired to outputs from 3D GCM as in \citeA{Vincendon2010water} and \citeA{Williams2022}.

In this context, the implementation of sub-grid scale slope microclimates in Mars GCMs is now required to better understand the present and past surface/subsurface processes that shape the Martian environment we know today. This problem is similar to the one faced by Earth climate modelers with the sub-grid scale diversity of topography,  soil, or vegetation on Earth.  They use several approaches to model the interaction between the land surface
heterogeneity with the overlying atmospheric layer which are known as \textit{Land Surface Parameterization}  \cite{Pitman2003, Flato2013,deVrese2016}.  Common approaches used, like the "mosaic" or "mixture" strategies, divides the mesh into homogeneous discrete sub-divisions where microclimates   evolve. Fluxes to the atmosphere are then computed by averaging all sub-grid scale values usually weighted by their respective cover fractions,  by using more complex parameterizations \cite{Giorgi1997},  or set to the values computed with the sub-division that has the largest cover fractions \cite{Dickinson1986}.

\begin{figure}[!ht]
    \centering
    \includegraphics[width = 1.2\textwidth]{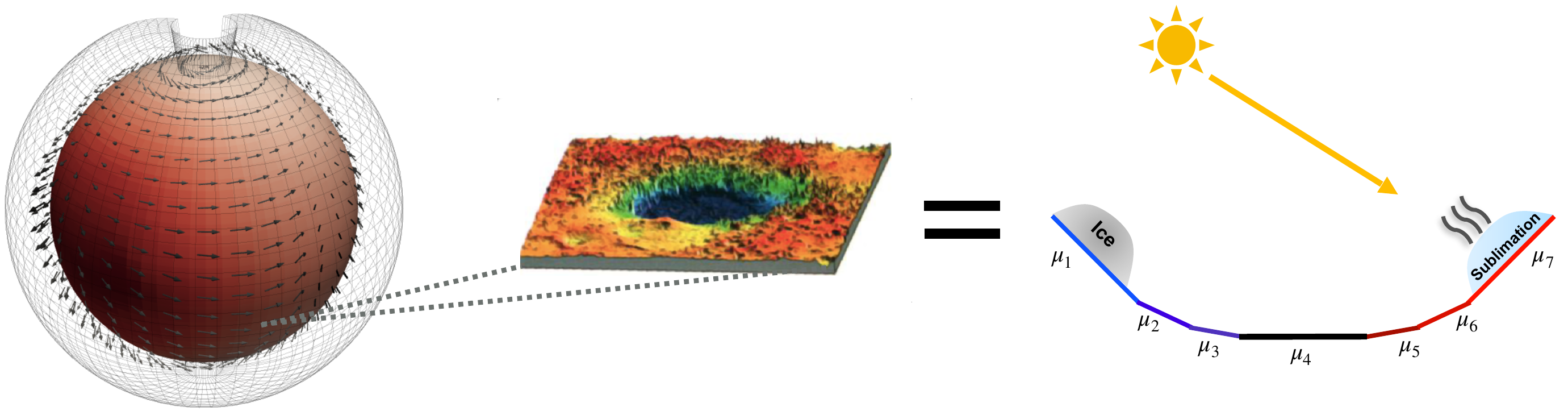}
    \caption{Schema of the sub-grid scale slope parameterization. Each coarse mesh of the global Mars PCM (left) is decomposed into sub-grid slopes (defined by characteristic slopes $\mu_i$) (right) or flat terrain, in order to represent the  slope distribution within the observed local topography (middle). These sub-grid terrains have their own microclimate (insolation, surface and sub-surface temperatures, condensation/sublimation of frosts, etc), and the interactions between the atmosphere and surface are made through averaged values over the mesh.  On the right, bluish surfaces mean colder slopes than flat surfaces (and warmer for reddish surfaces).  }
    \label{fig:cartoon_sub-gridslopes}
\end{figure}

We propose here to develop for Mars a similar sub-grid scale parameterization that accounts for topographic heterogeneity. The principle is illustrated in  Figure \ref{fig:cartoon_sub-gridslopes}. For each GCM mesh, we decompose the cell as a distribution of sloped terrains (defined by characteristic slopes) and a flat terrain. On each sub-grid terrain, we let the microclimates evolve so that the  slope-specific features  (e.g., condensation of volatiles, formation of glaciers, migration of subsurface ice, etc.) can be simulated. The portion of the atmosphere above the ground within the cell sees an average of these surface microclimates, and all sub-grid terrains see the same "shared atmosphere".

In section \ref{sec:ConstructionStatistique}, we present the construction of the sub-grid slope representation in the model. We then describe in section \ref{sec:ModeleImplementation} the implementation of slope microclimates in the Mars PCM. An illustration of the capability of the model is given in section \ref{sec:illustration}. The validation of these sub-grid scale microclimates is presented in section \ref{sec:Validation}. Perspectives on the use of this new model are presented in section \ref{sec:Perspectives}, and conclusions are drawn in section \ref{sec:Conclusions}.

\section{Representation of Slopes in the Mars PCM Meshes \label{sec:ConstructionStatistique}}
\begin{figure}[!ht]
    \centering
    \includegraphics[scale = 0.4]{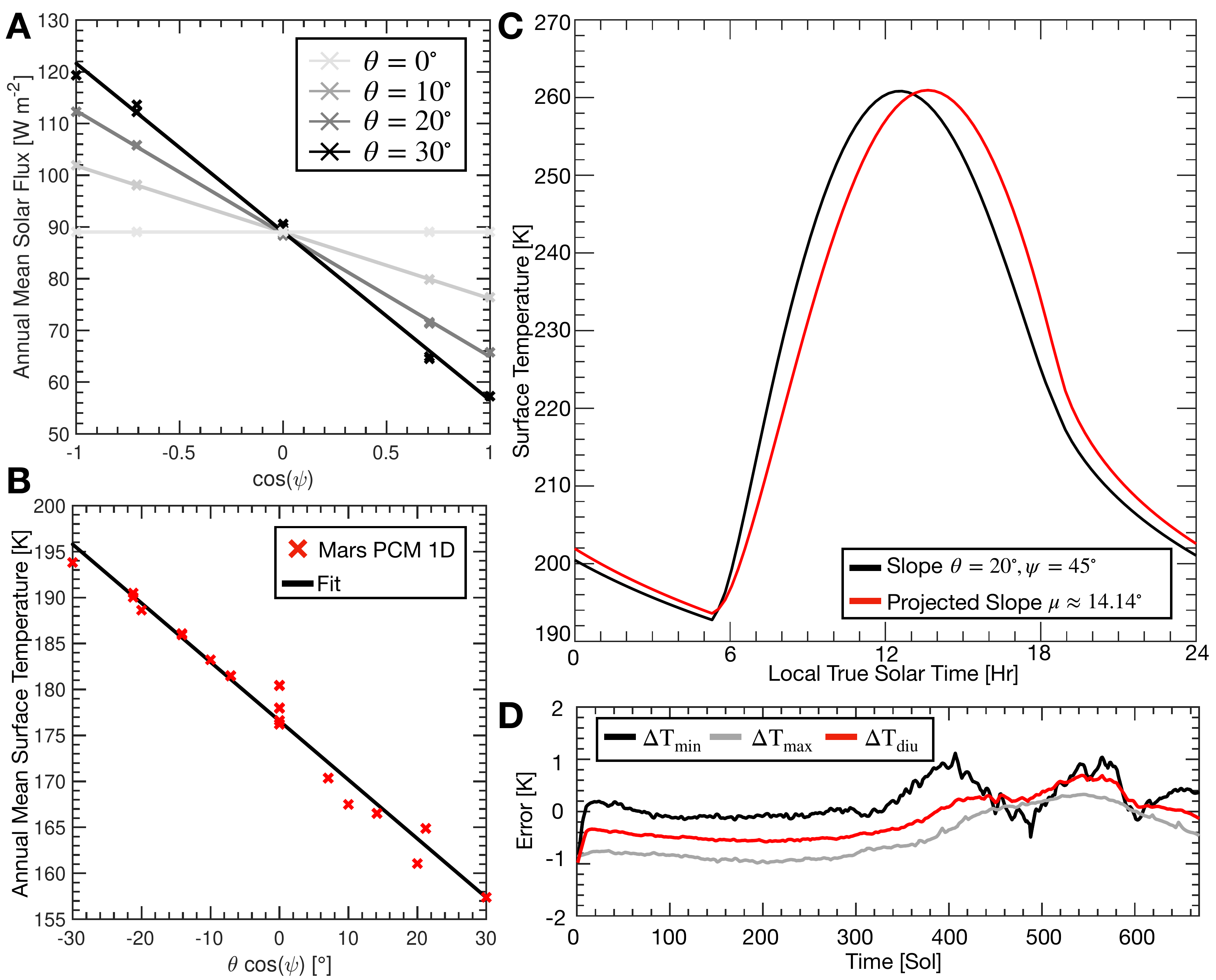}
        \caption{a) Annual mean solar flux for several slope angles $\theta$ and azimuths $\psi$ ($\psi$~=~~0\textdegree~corresponding to a slope oriented Northward) at a latitude of 70\textdegree N.  Crosses represent the outputs of the 1D Mars PCM, and the solid lines are the linear interpolation connecting these points.  Similar results are found at lower latitudes. b) Correlation between the mean annual surface temperature and the projected slope $\mu$. c) Diurnal evolution of the surface temperature for a slope with parameters $\theta~=~20$\textdegree~, $\psi = 45$\textdegree~(dark); and the corresponding projected slope $\mu$~$\approx$~14.14\textdegree~(red). The latitude is 30\textdegree~N.  d) Evolution of the error $ \Delta \rm{T}$ obtained during a Martian year between the minimum (black), maximum  (gray), and daily-averaged (red) surface temperature of the slope  $\theta = 20$\textdegree~, $\psi = 45$\textdegree~and the corresponding projected slope $\mu \approx$ 14.14\textdegree~. Similar results are found at other latitudes and with different slopes.}
    \label{fig:insolpente}
\end{figure}

The first step for the construction of our sub-grid slope parameterization is to determine a representative parameter of the surface inhomogeneity (i.e., the distribution of slopes within a mesh). We define the projected slope $\mu$ on the meridional direction as:
\begin{equation}
\mu = \theta \cos(\psi)  
\end{equation}
\noindent with $\theta$ the slope angle and $\psi$ the slope orientation. We demonstrate here that any given slope ($\theta$, $\psi$) can be thermally, on average, represented by a slope with a slope angle of
$\mid \mu \mid $ that is either North-facing if $\mu~>~0$, or South-facing if $\mu~<~0$.  $\mu$ will be thus the representative parameter for our distribution. To do so, we used the one-dimensional energy balance code derived from the Mars PCM \cite{Forget1999} and described in \citeA{Vincendon2010, Vincendon2010water}. In short, the 1D model computes the same physical  processes as in the 3D model and runs  with its local properties prescribed as inputs (slope angle, orientation, optical dept of the aerosols, etc.). The effects of large-scale meteorology or slope winds are not modeled in this 1D model. The 1D model uses the method described in \citeA{Spiga2008} to compute the radiative budget on slopes, taking into account scattering by aerosols and the thermal emission from adjacent flat surfaces. We ran the model for several slope angles and azimuths on several locations on Mars, using a nominal dust opacity profile derived by averaging the available observations of dust from Martian Year (MY) 24, 25, 26, 28, 29, 30, and 31 outside the global dust storm period \cite{MONTABONE2015} which is referred to as \textit{clim dust scenario} in the rest of the manuscript. Figure \ref{fig:insolpente}a illustrates satisfying linear correlations (R$^2 >$~0.95 in all cases) found between the parameter $\mu $ and the annual mean solar flux defined as the sum of the solar flux at infrared and visible wavelengths calculated by the model. Equivalent conclusions can be drawn using other simulations made at different latitudes. Similarly, a linear relationship (R$^2 >$~0.97) can be found between the projected slope and the surface temperature (Figure \ref{fig:insolpente}b). This linear relationship illustrates the fact  that for any given slope ($\theta$, $\psi$), the radiative energy budget can be, on average, represented by a slope with a slope angle of $\mid \mu \mid$ that is either North-facing if $\mu~>~0$, or South-facing if $\mu~<~0$. 

We test this assumption by computing several surface temperatures for a given slope ($\theta, \psi$) and the corresponding North/South-facing slope with a slope angle  $\mid \mu \mid$ at different latitudes. An illustration is given in Figure \ref{fig:insolpente}c for a  diurnal cycle and confirms that  the North/South slope emulator has the same thermal behavior as the initial slope ($\theta,\psi$). The difference between the two surface temperatures  is caused by the difference in the slope's azimuths, which create a phase shift that evolves during the sol and the position of the sun in the sky. Apart from this phase shift, the minimum, maximum, and daily-mean surface temperatures are consistent (relative error lower than 1\%) throughout the year (Figure \ref{fig:insolpente}d). The instantaneous error, created by the time lag between the two temperatures, barely exceeds 5\% of the instantaneous surface temperature and is neglected. Therefore, this confirms that any sub-grid scale slopes in the Mars PCM can be either represented by pole-facing slopes or equator-facing slopes when one wishes to study the behavior of ice or any process sensitive to the diurnal or seasonal variation of surface and subsurface temperatures. Yet, our model might be limited when studying processes on East/West slopes due to the time lag between  the North/South slope and the initial slope ($\theta,\psi$). One other approach used by \citeA{Williams2008} is to consider an equivalent latitude and an adjusted longitude computed from $\theta$,$\psi$ to correct this time shift. However,  this approach was more difficult to implement in the current Mars PCM without increasing significantly the computation time as it requires recomputing  the radiative transfer at each equivalent latitude and an adjusted longitude. Thus,  this approach  has not been adopted for our parameterization. Furthermore, in this 1D experiment, we do not consider the possible heterogeneity of albedo, thermal inertia, or the possible influence of slope winds. These effects are discussed later in section \ref{ssec:discusshetereogenenity}, \ref{ssec:discussslopewind}.

\begin{figure}[!ht]
    \centering
    \includegraphics[scale = 0.4]{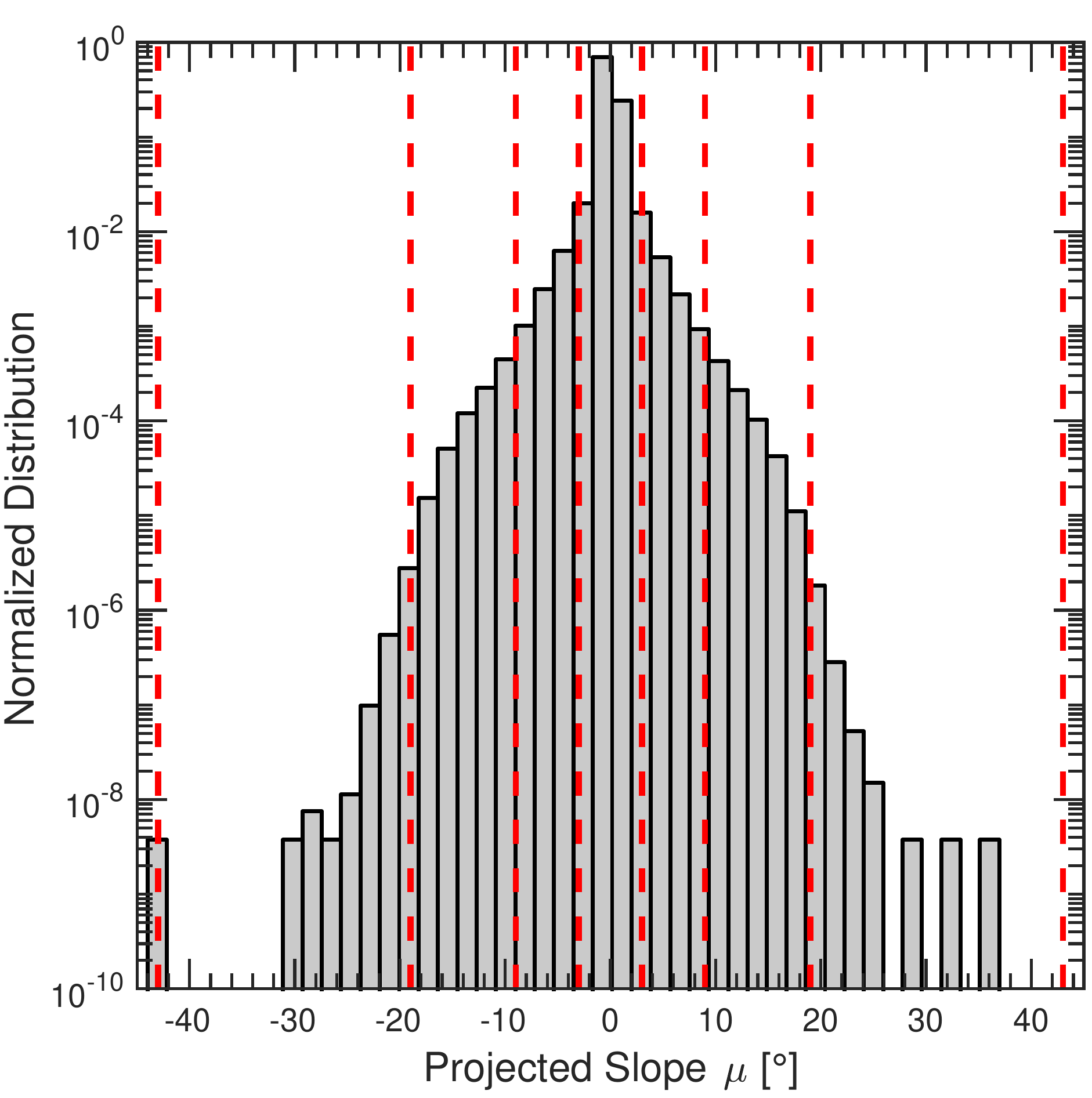}
    \caption{Distribution of the projected slopes $\mu$ computed with MOLA 1/64\textdegree~dataset. The distribution has been normalized by dividing each count per pixel by the total number of pixels ($\sim 3\times 10^8$). Red dashed lines represent the boundaries of the slope classes (see Table \ref{Table:Statistique} for definitions).
}
    \label{fig:distribution_mu}
\end{figure}

We then need to find how to discretize the north-south projected slopes $\mu$ to decompose our mesh into a limited subset of representative sub-grid scale slopes (Figure \ref{fig:cartoon_sub-gridslopes}). To do so, we studied the distribution of $\mu$  computed with the Mars Orbiter Laser Altimeter \cite<MOLA,>{Smith2001} data that have been widely used to study the Martian topography and slope distribution \cite<e.g.,>{Smith2001, Kreslavsky2000, Aharonson2003}. We used the 1/64\textdegree~dataset corresponding to a resolution of nearly $\sim$~1~km at the equator which corresponds to the typical length of key geomorphological features observed at the surface of Mars  \cite<e.g.,>{Head2003, Kreslavsky2011, Hobbs2013}. Note that the MOLA resolution used depends on the user's desired resolution for the sub-grid surface, and that lower or higher resolutions can be used.  The MOLA data used are in a simple cylindrical format, and we have not corrected them for Mars' non-sphericity, assuming it has only a minor effect. Given the two-dimensional topographical field $h(x, y)$ where $x$ and $y$ are the distance in meters respectively in the longitude and the latitude direction, and $h$ the altitude in meters at coordinates $(x,y)$, the slope angle $\theta$ and the slope orientation $\psi$  can be computed with Eq. \ref{eq:Eq.1} and \ref{eq:Eq.2} \cite{Spiga2008}:

\begin{equation}
    \tan \theta = \sqrt{\left( \frac{\partial h}{\partial x} \right)^2 + \left( \frac{\partial h}{\partial y} \right)^2}
    \label{eq:Eq.1}
\end{equation}

\begin{equation}
    \tan \psi = \frac{ \partial_y h}{\partial_x h}
    \label{eq:Eq.2}
\end{equation}

\noindent where $\partial_y h = \frac{\partial h}{\partial y}$, $\partial_x h = \frac{\partial h}{\partial x}$. Derivatives are calculated using centered differences with  neighboring pixels. The projected slope $\mu = \theta \cos(\psi)$ is then computed. The distribution of $\mu$ is presented in Figure \ref{fig:distribution_mu}. A complete analysis of the Martian slope distribution is out of the scope of this paper, but the results obtained here (predominance of flat surfaces on Mars, the limited number of steep slopes, etc.) are consistent with dedicated studies using the same dataset and similar resolution \cite{Kreslavsky1999, Kreslavsky2000}. The distribution of slopes obtained is sensitive to the resolution of the MOLA dataset used. As noted in  \citeA{Aharonson2006}, a high-resolution topographic dataset promotes steep slopes at the expanse of flat surfaces. Yet, at a resolution of hundreds of meters, their results (see their Figure 6) are consistent with ours, i.e., flat surfaces predominate over slopes. At very low scales (resolution $\leq$~10~m), where steep slopes are significant contributors, the Mars PCM  physics and parameterizations are not adapted and would be in the domain of the Large-Eddy Simulations.

We then defined seven slope classes (delimited by the red lines in Figure \ref{fig:distribution_mu} and defined in Table \ref{Table:Statistique}) that are representative of the projected slope distribution. Each mesh of the Mars PCM is  divided into 7 sub-grid scale slopes that are either north-facing or south-facing, defined by a characteristic $\mu$ presented in Table \ref{Table:Statistique}.  We assume that this partition with seven sub-grid surfaces (1 for the flat one, and 6 for the slopes) allows us to represent the distribution of slopes within the meshes. We could have added more classes for the decomposition but this would have increased the computation time too much without much scientific gain. Also, we cannot decrease the number of classes too much because we would lose a realistic representation of the distribution within the mesh (i.e., we would have a flat terrain and only extreme slopes).  The characteristic $\mu$ proposed in Table \ref{Table:Statistique} are computed as the means of the endpoints of the slope classes (except for the class 1 and 7 where it is set to 30\textdegree), rather than an average of $\mu$ weighted by the distribution within the class. The typical slope used to represent a class is exaggerated (by $\sim$1\textdegree~for classes 3-5; $\sim$3\textdegree~for classes 2-6 for instance, $\sim$5\textdegree~for classes 1-7).  This choice was motivated by future studies which required modeling steep ($\geq $25\textdegree) slopes (see for instance sections \ref{sec:Validation}, \ref{sec:Perspectives}).

\begin{table}[!ht]
\centering
\begin{tabular}{cccll}
\cline{1-3}
\multicolumn{1}{|c|}{Class} & \multicolumn{1}{c|}{Definition}                                     & \multicolumn{1}{c|}{~Characteristic $\mu$} &  &  \\ \cline{1-3}
\multicolumn{1}{|c|}{1}                    & \multicolumn{1}{c|}{  -43\textdegree~$\leq \mu \leq$ -19\textdegree~} & \multicolumn{1}{c|}{$\mu_1 = -30$\textdegree~}                   &  &  \\ \cline{1-3}
\multicolumn{1}{|c|}{2}                    & \multicolumn{1}{c|}{ -19\textdegree~$\leq \mu \leq$ -9\textdegree~}  & \multicolumn{1}{c|}{$\mu_2 = -14$\textdegree~}                   &  &  \\ \cline{1-3}
\multicolumn{1}{|c|}{3}                    & \multicolumn{1}{c|}{ -9\textdegree~$\leq \mu \leq$ -3\textdegree~}   & \multicolumn{1}{c|}{$\mu_3 = -6$\textdegree~}                    &  &  \\ \cline{1-3}
\multicolumn{1}{|c|}{4}                    & \multicolumn{1}{c|}{ -3\textdegree~$\leq \mu \leq$ 3\textdegree~}    & \multicolumn{1}{c|}{$\mu_4 = 0$\textdegree~}                     &  &  \\ \cline{1-3}
\multicolumn{1}{|c|}{5}                    & \multicolumn{1}{c|}{  3\textdegree~$\leq \mu \leq$ 9\textdegree~}     & \multicolumn{1}{c|}{$\mu_5 = 6$\textdegree~}                     &  &  \\ \cline{1-3}
\multicolumn{1}{|c|}{6}                    & \multicolumn{1}{c|}{  9\textdegree~$\leq \mu \leq$ 19\textdegree~}    & \multicolumn{1}{c|}{$\mu_6 = 14$\textdegree~}                    &  &  \\ \cline{1-3}
\multicolumn{1}{|c|}{7}                    & \multicolumn{1}{c|}{  19\textdegree~$\leq \mu \leq$ 43\textdegree~}   & \multicolumn{1}{c|}{$\mu_7 = 30$\textdegree~}                    &  &  \\ \cline{1-3}
\multicolumn{1}{l}{}                       & \multicolumn{1}{l}{}                                             & \multicolumn{1}{l}{}                                &  & 
\end{tabular}
    \caption{Description of the under-mesh slope parameterization with the slope classes, their definition, and the characteristic projected slope $\mu$ associated.}
    \label{Table:Statistique}
\end{table}

\begin{figure}[!ht]
    \centering
    \includegraphics[scale = 0.45]{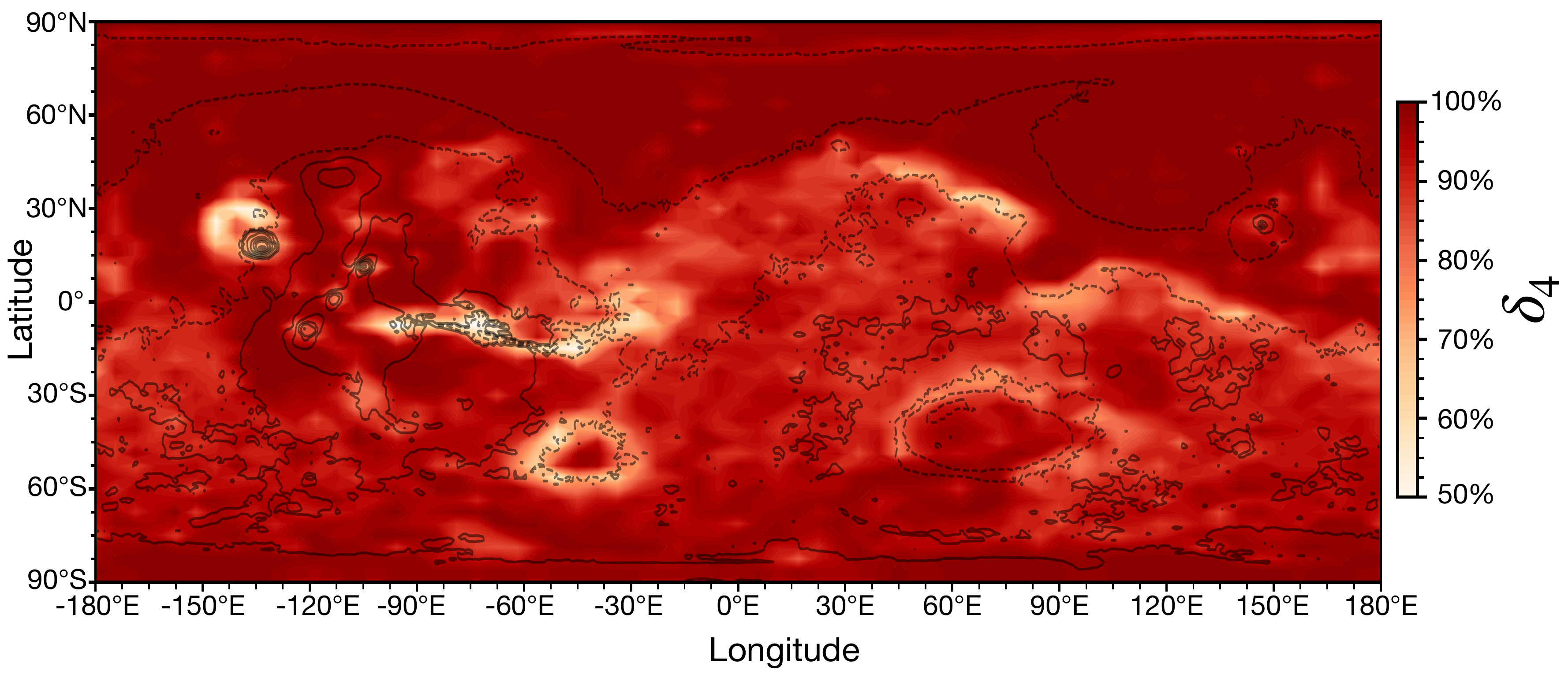}
    \caption{Cover fraction of the PCM meshes for projected slopes $\mu_4$~($\delta_4$). The redder the cell, the flatter the terrains in the mesh. Black dashed lines are MOLA topographic contours. 
}
    \label{fig:mapslope}
\end{figure}

For each mesh of the Mars PCM and each characteristic slope $\mu_i$, we can compute the percentage of the area of the cell occupied by slopes that are in a class $i$ (this percentage is noted $\delta_i$ hereinafter and called \textit{cover fraction}). In this computation, the variable size of the MOLA pixels within a Mars PCM grid because of the latitude is considered.  Here, and for the rest of the manuscript (if not explicitly stated), the resolution of the Mars PCM grid was  5.625\textdegree~in longitude, and 3.75\textdegree~in latitude (i.e., 330~km $\times$ 220~km, a standard resolution for the PCM for which the high resolution is 60~km $\times$ 60~km, i.e., 1\textdegree~$\times$~1\textdegree). Results for class 4, i.e., for quasi-flat terrains, are presented in Figure \ref{fig:mapslope}. The distribution of the flat sub-grid surface  obtained (Figure \ref{fig:mapslope}) is consistent with what is presented in the literature \cite<e.g., Figure 6 of >{Aharonson2006}. Most of the sloped terrains are logically found in Valles Marineris, Hellas Planitia, Elysium Planitia, and the North-South dichotomy.   It should be noted that at this spatial resolution,  the predominant sub-grid surface (i.e., the one with the largest cover fraction) is the flat terrain. To avoid numerical artifacts, we assume that the whole mesh is flat when the area occupied by the slopes $\mu_{1,2,3,5,6,7}$ is smaller than 0.001\% of the total mesh area.

Another important quantity for our parameterization is the shape of the distribution of the sub-grid slopes within a mesh. Are the sub-grid slopes uniformly distributed or on the contrary, a  sub-slope is largely predominant? We calculated for each mesh the kurtosis of the real (and not discretized as in  Table \ref{Table:Statistique}) distribution of the sub-grid slopes computed from the MOLA dataset. The kurtosis is a mathematical parameter that measures the tailedness of the distribution. To simplify, a high kurtosis suggests that the distribution is highly concentrated around its mean, while a low kurtosis suggests a rather uniform distribution. All the calculated kurtosis are high, even for cells where the flat sub-grid surface is less dominant. The distribution of the sub-grid slopes is very narrow and centered on the flat sub-slope. We also tested these properties with a high-resolution GCM grid (1\textdegree$\times$1\textdegree). In that case, for some meshes, the predominant sub-grid surface is a sloped one (e.g., in Valles Marineris), and the distribution is still tight and centered on this slope. Only 1\% of the GCM grid does not have a clear peak in the sub-grid slope distribution.  The consequences of the properties of the distribution are discussed in the next section.

\section{Sub-Grid Scale Microclimates Modelling \label{sec:ModeleImplementation}}

\begin{figure}[!ht]
    \centering
    \includegraphics[scale = 0.13]{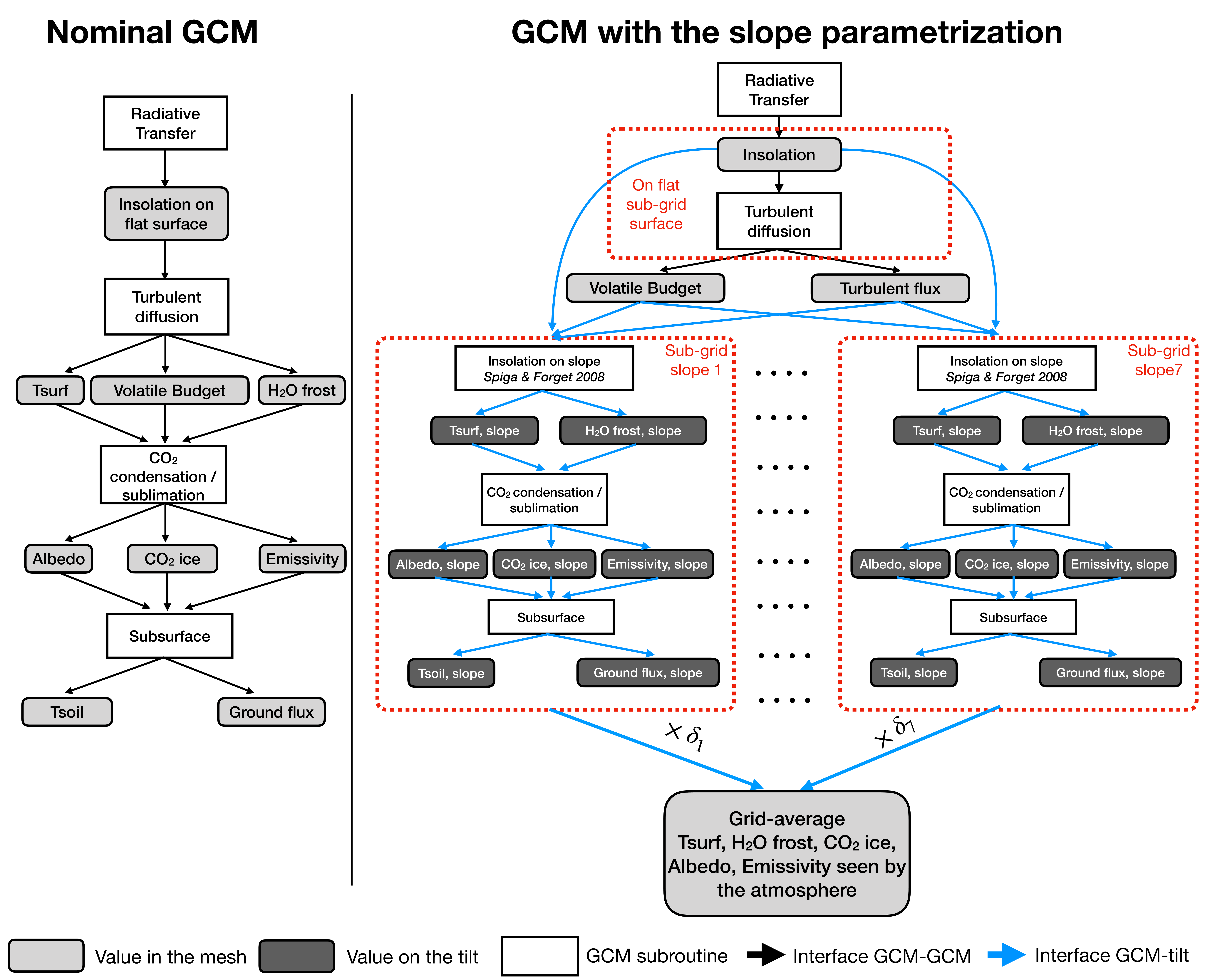}
    \caption{ Description of the implementation of the sub-grid scale slope microclimates in the Mars PCM.  The left part presents the architecture of the Mars PCM without the contribution of the sub-grid slopes while the right part presents the new architecture of the model, with the parameterization.  Processes that only involved the atmosphere and not the surface in the Mars PCM have not been represented in the figure to clarify the diagram. $T_{\rm{surf}}$ corresponds to the surface temperature while $T_{\rm{soil}}$ refers to the subsurface soil temperature.
}
    \label{fig:cartoon_microclimate}
\end{figure}

We propose the following parameterization to simulate slope microclimates,  inspired by the  "mosaic" and "mixture" strategies in Earth models \cite{deVrese2016}. On each binned slope, we calculate the radiative energy budget at the mesh center. Section \ref{sec:ConstructionStatistique} showed us that, at the given spatial resolution,  the flat sub-grid surface strongly dominates. One can view the Martian topography within a mesh as a predominantly flat terrain with some sporadic slopes (where a significant amount of volatiles can condense, however).  Turbulent exchanges mostly take place with the flat terrain, the slopes making only negligible contributions to the mixing.   We thus assume that  turbulent exchanges are solved with the predominant sub-grid surface.  We compute and let evolve independently on each sub-grid slope the surface and subsurface temperatures, the volatile budget, and ground properties. We assume that these sub-grid slopes are  independent and share the same atmosphere. The atmosphere only sees an averaged value of these surface fields, weighted by their cover fraction. The implementation of this strategy is detailed in Figure \ref{fig:cartoon_microclimate}.

\subsection{Energy budget on a slope}

The evolution of the  surface temperature is  computed based on the energy budget at the surface:

\begin{equation}
  \rho  c_s \frac{\partial T_{\rm{surf,slope}}}{\partial t} = (1-A_{\rm{slope}}) F_{\rm{rad,sw}}(t) + \epsilon_{\rm{slope}}  F_{\rm{rad,lw}}(t) +  F_{\rm{ground}}(t) +  F_{\rm{atm}}(t) + \sum_{i}L_i\frac{\partial m_i}{\partial t}   - \epsilon_{\rm{slope}} \sigma T_{\rm{surf,slope}}^4(t)
  \label{Eq:energysurflayer}
\end{equation}

\noindent where the left member of the equation is the energy of the surface layer, with $\rho$ is the  density of the ground (kg~m$^{-3}$),   c$_s$ is the "surface layer" heat capacity per unit area (J~m~kg$^{-1}$~K$^{-1}$) and is related to the thickness of the first soil layer,  $T_{\rm{surf,slope}}$ the sub-grid slope temperature (K), $A_{\rm{slope}}$ (unitless) the albedo of the surface, $F_{\rm{rad,sw}}$ the radiative flux at visible wavelengths (W~m$^{-2}$), $\epsilon_{\rm{slope}}$~(unitless) is the surface emissivity (assumed to be equal to the absorptivity), $F_{\rm{rad,lw}}$ the radiative flux at infrared  wavelengths (W~m$^{-2}$), $F_{\rm{ground}}$ the soil heat flux due to heat conduction process (W~m$^{-2}$), $F_{\rm{atm}}$ the sensible heat flux, $\sum_{i}L_i\frac{\partial m_i}{\partial t} $ the latent heat flux due to the condensation/sublimation of a volatile with a latent heat $L_i$ (J~kg$^{-1}$) and $ \epsilon_{\rm{slope}} \sigma T_{\rm{surf,slope}}^4$ the radiative cooling of the surface with $\sigma$~=~5.67$\times 10^{-8}$ W~m$^{-2}$~K$^{-4}$ the Stefan-Boltzmann constant. We describe below the computation of each term.

The resolution of Eq. \ref{Eq:energysurflayer} is made using an implicit scheme and the soil heat conduction is solved using a  finite volume approach to the problem. Surface temperature
is technically linked to atmospheric and ground processes which are
coupled. Following the strategy of \citeA{Hourdin1995}, the resolution is made as follows:  the internal energy of the surface layer, and thus the surface temperature, is updated at time $t$ using Eq. \ref{Eq:energysurflayer}. Temperatures of  the subsurface layers are then updated by solving the thermal conduction. Once these temperatures are computed, the soil heat flux is computed and given as an input of Eq. \ref{Eq:energysurflayer} for the next time step.

\subsubsection{Radiative Fluxes \label{ssec:radtransfer}}
The  computation of radiative fluxes accounts for \CO gas infrared absorption/emission \cite{Hourdin1992,Forget1999,Dufresne2005},  dust absorption, emission,  diffusion in the visible and infrared \cite{Forget1998,Forget1999,Madeleine2011}, and the radiative effect of water ice clouds \cite{Madeleine2012}. Dust opacity is set based on the maps from \citeA{MONTABONE2015} and we used in the following the \textit{clim dust scenario}.

The total irradiance computed with the current method must be adapted when considering a sloped terrain. $F_{\rm{rad,sw}}$, which accounts for direct incoming flux from the sun, scattered flux by dust in the atmosphere, and the reflected flux from the neighboring terrains is computed with the parameterization from \citeA{Spiga2008}. For thermal infrared radiation, we assume that 
the atmospheric thermal radiation is isotropic and we note $F_{\rm{IR,flat}}$~(W~m$^{-2}$) the atmospheric incident thermal infrared flux on a horizontal surface computed by the PCM.  Here, for a given sub-grid slope, we assume that the rest of the mesh is flat to be consistent with \citeA{Spiga2008}. The sky-view factor  $\sigma_s$ (unitless), which  quantifies the proportion of the sky in the half hemisphere seen by the slope that is not obstructed by the surrounding terrain, is defined as in \citeA{Spiga2008}:
\begin{equation}
\sigma_s    = \frac{1+\cos(\mid \mu \mid )}{2}
\end{equation}
\noindent where $\mid \mu \mid$ refer to the absolute value of the projected slope $\mu$.

Based on geometrical considerations, the thermal infrared radiation for a sloped surface  $F_{\rm{IR,slope}}$ can be written as 
\begin{equation}
     F_{\rm{rad,lw}} = \sigma_s  F_{\rm{rad,lw,flat}} + (1-\sigma_s)\epsilon_{\rm{flat}}\sigma T_{\rm{surf,flat}}^4
     \label{eq:eqFir}
\end{equation}

\noindent where $\epsilon_{\rm{flat}}$ (unitless) is the emissivity sub-grid flat terrain,  $T_{\rm{surf,flat}}$ (K) its temperature. Here, we neglect the possible influence of the atmosphere (both in terms of absorption and diffusion)  
on the emission from the flat surface.

\subsubsection{Sensible and Latent Heat Flux \label{ssec:latent} }

The sensible heat flux describes the energy exchange between the atmosphere and
surface due to molecular conduction and turbulence. Such flux can be written as \cite{Forget1999}:
\begin{equation}
    F_{\rm{atm}} = \rho C_H U c_{\rm{air}}(T_{\rm{atm, 1}} - T_{\rm{surf}})
    \label{eq:turb}
\end{equation}
\noindent where  $U$ (m~s$^{-1}$) is the wind velocity obtained by combining
the large-scale (synoptic) wind near the surface with a gustiness wind induced by buoyancy \cite{Colaitis2013}, $c_{\rm{air}}$ (J~kg$^{-1}$~K$^{-1}$) is the air specific heat capacity at constant pressure, $T_{\rm{atm, 1}} $ (K) the temperature of the atmosphere in the 1$^{st}$ vertical layer $z_1$ of the GCM (about 4~m above the surface) and $C_H$ (unitless) is a heat transfer coefficient  given by \cite{Colaitis2013}:

\begin{equation}
    C_H = f_h(Ri) \left( \frac{\kappa^2}{\ln{\frac{z_1}{z_0}}  \ln{\frac{z_1}{z_{0T}}}}\right)
\end{equation}

\noindent where $f_h(Ri)$ is a function of the Richardson number $Ri$,  $\kappa$~(unitless) is the von Karma constant set to 0.4; $z_0$ (m) is the aerodynamic roughness coefficient extracted from \citeA{Hebrard2012},  $z_{0T}$ (m) the thermal roughness length. The computations of $f_h(Ri)$ and $z_{0T}$  are detailed in \citeA{Colaitis2013}.  $U$ and $T_{\rm{atm, 1}} $ are computed when considering the predominant sub-grid terrain.

For each sub-grid surface, water ice deposition and sublimation are computed following \citeA{Navarro2014}:
\begin{equation}
    \frac{\partial m_{w}}{\partial t}= \rho C_H U (q_w - q_{sat}(T_{\rm{surf}}))
    \label{eq:sublimwaterice}
\end{equation}
\noindent where $m_w$~(kg~m$^{-2}$) is the mass of \HHO frost, normalized by unit area, $q_{w}$ is the mass mixing ratio of water vapor in the first layer, $q_{sat}$ is the saturation mass mixing ratio at the surface temperature \cite<see Eq. (1) of >{Pal2019}. The latent heat flux is then computed by multiplying by $ \frac{\partial m_{w}}{\partial t}$ with the latent heat of sublimation of water at $T_{\rm{surf}}$, computed from a 2nd-order polynomial fit of \citeA{FEISTEL200736}'s data. 

In the model, convection near the surface promotes water sublimation through the gustiness wind induced by buoyancy \cite{Colaitis2013} and the stability function $f_h$. However, this convection is assumed to be dry, induced by the thermal contrast between the surface and the atmosphere. We, therefore, neglect convection induced by the molecular gradient of the air linked to the presence of water near the surface \cite<e.g.,>{Ingersoll1970}. We assume that the vapor pressure of water at the surface is too low to affect the molecular weight of the air and that turbulent mixing at the surface should limit the impact of this free convection \cite{Ivanov2000}. However, this assumption can be questioned in the case of  frost with warm temperatures close to the melting point \cite{Schorghofer2020}. However, the model predicts water ice with a  temperature $\leq$~220~K, which legitimizes our approximation. However, further work is needed to include these effects in future studies, particularly in the case of ice stability at high obliquity, where melting is expected \cite{Costard2002}. This requires a reparameterization of surface fluxes and a wet convection model that are beyond the scope of the paper. Nevertheless, the effect of the water latent heat term is generally neglected on Mars as it is often reduced to $\sim$1-2~W~m$^{-2}$ today \cite<e.g.,>{Richardson2002, Martinez2014}; although it can become significant at higher obliquity \cite{Naar2021}.

\CO deposition and sublimation computations and the latent heat associated are made following \citeA{Forget1998snow}. For ground condensation, as \CO is abundant in the atmosphere if the surface temperature on a sub-grid slope predicted by radiative and conductive balance $T_0$ (i.e., Eq. \ref{Eq:energysurflayer} solved without any condensation/sublimation of \CO~)   falls below the condensation temperature  at surface pressure $T_{\rm{cond}}$ (given by the Clapeyron's law using the partial pressure of \CO, \citeA{Jammes1992}), the amount of \CO condensing $\delta m$ (kg~m$^{-2})$ is:

\begin{equation}
    \delta m = \frac{\rho c_s}{L_{CO_2}+ c_{\rm{air}}(T_{\rm{atm, 1}} - T_{\rm{cond}})}\left(T_{\rm{cond}} - T_0 \right)
\end{equation}

\noindent where $L_{CO_2}$ (J~kg$^{-1}$) is the latent heat of \CO. This expression is almost similar to the one of \citeA{Forget1998snow} except we add the term  $c_{\rm{air}}(T_{\rm{atm, 1}} - T_{\rm{cond}})$  which corresponds to the extra heat brought by the atmosphere when cooled to the condensation temperature $T_{\rm{cond}} $ just above the surface. Even if the temperature $T_{\rm{atm, 1}}$ is cooled by Eq. \ref{eq:turb} and radiative processes, in the case of a pole-facing slope,  the lower atmosphere lying just above the surface can be significantly warmer than the surface so this term can be significant. Typically, $T_{\rm{atm, 1}}$ can be 15~K higher than on frosted pole-facing slopes. As such, the extra heat brought by the atmosphere when cooled can reach nearly 2\% of the latent heat.   In the case of atmospheric precipitation (\CO snow, but also other volatile deposition), we assume a uniform deposition on the whole mesh. Once $\delta m$ is known, it is injected in Eq. \ref{Eq:energysurflayer} which will give, after integration, $T_{\rm{surf}}$ to be $T_{\rm{cond}}$ if there is \CO ice on the surface.

\subsubsection{Conduction in the soil and subsurface water ice\label{ssec:groundice}}
The computation of the ground flux is based on the  1D layer soil model originally described by \citeA{Hourdin1993} but rewritten for a physical vertical grid (in meters). The current Mars PCM solves the conduction in the soil from 2~$\mu$m to 25~m, corresponding respectively to the minimum diurnal skin depth for a thermal inertia of 30 J~m$^{-2}$~K$^{-1}$~s$^{-1/2}$ and the maximum annual skin depth for a thermal inertia of 2000 J~m$^{-2}$~K$^{-1}$~s$^{-1/2}$, range of thermal inertia observed on Mars \cite{MELLON2000}.  Optical properties (albedo, emissivity) of the surface are set as described in Table \ref{Table:PCMinput}  based on the Thermal Emission Spectrometer (TES) measurements \cite{ Christensen2001}. We assume that all the frost-free sub-grid surfaces have the same optical properties. Compared to the former version of \citeA{Hourdin1993}, the current Mars PCM  enables variations in the thermal properties (volumetric specific heat, thermal conductivity) with depth so that the effect of  subsurface water ice can be included in the GCM. The high and mid-latitude subsurface ice reservoirs have a  significant impact on the Martian climate. Because of their high thermal inertia, they store large amounts of heat during summer and restore them during winter. It thus significantly affects the  \CO cycle through the ice deposition/sublimation \cite{Haberle2008} and can prevent/reduce the condensation of \CO on pole-facing slopes \cite<see > [for a complete discussion on the effect of subsurface ice on slope energy budget]{Vincendon2010,KhullerGullies}.  In the current model that does not consider sub-grid slopes, maps of the ice table are created using measurements by the Mars Odyssey Neutron Spectrometer (as part of GRS) of the abundance of hydrogen near the surface \cite{Feldman2002, Feldman2004} and an inversion model of the effective depth of the water ice table \cite{DIEZ2008}. 

Few observational constraints on the ice table under a slope exist \cite<e.g.,>{Khuller2021ice} because of 1) the resolution of the instruments (more than 500 km for the spectrometers \cite{Saunders2004}); 2) the paucity of high-resolution observations allowing a fine determination of the ice table on the slopes \cite{Bandfield2007, Vincendon2010, Piqueux2019}. To take into account the thermal effect of the ice table on the slopes, we must model their distribution. To do so, we used the theory proposed by \citeA{Leighton1966, Mellon2004, Schorghofer2005, Aharonson2006}:  a subsurface ice (either massive ice or pore-filling)  at a depth \textit{z} is stable if: 

\begin{equation}
    \overline{\frac{p_{\rm{vap,surf}}}{T_{\rm{surf}}}} \geq \overline{\frac{p_{\rm{sv}}(T_{\rm{soil}}(z))}{T_{\rm{soil}}(z)}}
\end{equation}
\noindent where overbars indicate time averages over a complete Martian year,  $p_{\rm{vap, surf}}$~(Pa) is the water vapor pressure at the surface, $p_{\rm{sv}} $~(Pa) is the saturation water vapor pressure over water ice which is a function of the temperature \cite{Murphy2005}. \citeA{Mellon1993, Schorghofer2005, Aharonson2006} used radiative equilibrium with fixed humidity, based on the values retrieved by the TES \cite{Smith2002} and do not simulate the complete water cycle. Recent sensitivity studies have shown that subsurface water ice stability is sensitive to atmospheric conditions \cite{SONG2023} as well as the presence of surface water frost \cite{Hagedorn2007, McKay2009, Williams2015}. Here, with the  Mars PCM, we can compute more accurately the water cycle which has been validated through TES data \cite{Navarro2014, Naar2021} and thus the vapor density at the surface considering all of these effects. We set $p_{\rm{vap,surf}}$ to the near-surface vapor pressure given by the GCM if the surface is not ice-covered, $p_{\rm{sv,surf}}(T_{\rm{surf}})$  if it is ice-covered. When ice is stable at depth $z_{ice}$, we set the thermal inertia  of this layer and those below to 1,600~J~m$^{-2}$~K$^{-1}$~s$^{-1/2}$, 
a mid-value between completely pore-filled ice \cite<thermal inertia of $\sim$1200~W~m$^{-2}$~K$^{-1}$; Eq. (26) of >{Siegler2012} and massive pure ice \cite<thermal inertia of $\sim$2050~W~m$^{-2}$~K$^{-1}$ at 180~K;>{Hobbs1974}.

As water ice enhances conduction and thus the propagation of thermal waves to depth, it attenuates shallow depth temperature fluctuations, which decrease the mean vapor density and promotes the stability of subsurface water ice. Hence, the algorithm is iterated until an equilibrium is reached for the ice table depth. With this approach, we  simulate for the first time  the distribution of the ice table underneath slopes but also  flat surfaces with a complete GCM for the atmosphere and the (sub)surface. The results will be presented in a dedicated paper \cite{Lange2023ice}.

 \begin{table}[h]
 \begin{center}
     
 \centering Map Inputs
 \begin{tabular}{p{0.15\textwidth}p{0.2\textwidth}p{0.2\textwidth}p{0.3\textwidth}p{0.2\textwidth}} \hline
Input & Source & Thinnest Spatial Resolution &	Data Uncertainty & Resolution used \\
\hline
Topography	& MOLA \cite{Smith1999} &500 m	& $\sim$1–3~m \cite{Neumann2001} & 1 km	  \\
Thermal Inertia	& TES	\cite{Putzig2005}& 3 km	 & ~8\% of the thermal inertia value	\cite{Putzig2005}& PCM resolution	\\
Albedo – bare ground &	TES	\cite{Putzig2005} & 3 km	& ~0.5\% of the albedo value	 \cite{Putzig2005}& PCM resolution	\\

Surface Roughness	& Derived from TES rock abundance \cite{Hebrard2012}	& 7 km& 	Between 10$^{-2}$ to 10$^{-4}$ m \cite{Hebrard2012} & 	PCM resolution \\
\hline
 \end{tabular}
 \centering Constant inputs
\begin{tabular}{p{0.25\textwidth}p{0.6\textwidth}p{0.25\textwidth}} \hline
Input & Range of value observed & Value used in this study \\
\hline
Albedo - \HHO ice & 0.25-0.6, depending on dust contamination and grain size (Langevin 2005) & 0.33 \\
Albedo - \CO ice & 0.25-0.65 \cite{Titus2001,Langevin2007} & 0.65 \\
Emissivity – bare ground&	0.75-1 \cite{Bandfield2003}& 0.95	\\
Emissivity – H2O ice	& 0.95-1 \cite{Vincendon2010water} & 1	\\
Emissivity – CO2 ice	& 0.8-1 \cite{Titus2001,Piqueux2016} & Computed after \citeA{Forget1998snow}\\
\hline
 \end{tabular}

 \end{center}
\caption{Values of the different inputs for the PCM simulation. 3D field inputs are presented in the upper table, while 1D inputs are given in the lower table. TES refers to the Thermal Emission Spectrometer \cite{Christensen2001}. When needed, input maps are degraded to the PCM resolution using bilinear interpolation. Albedo refers here to broadband albedo.}
\label{Table:PCMinput}
 \end{table}

\subsection{Interaction between sub-grid surfaces and the atmosphere \label{ssec:gridbox}}
Once the microclimates have been computed and sub-grid surface properties updated, we need to communicate this information to the atmosphere by updating the mesh values. Following Earth model conventions \cite{deVrese2016}, grid parameters are determined by averaging all sub-grid parameter values weighted by their respective cover fractions. In practice, we assume that the grid box albedo and emissivity can be computed with :
\begin{equation}
     X = \sum_{i=1}^7 X_i \delta_i
     \label{eq:opticalproperties}
\end{equation}
\noindent where $X$ refers to the grid parameter, $X_i$ the sub-grid parameter of slope $i$, $\delta_i$ the cover fraction.

In the model, condensed volatiles at the surface (i.e., seasonal frost or perennial ice) are represented with surface mass density (kg~m$^{-2}$). However, the area of the sloped mesh grid is not the same if it is sloped or flat, assuming a similar latitude/longitude boundary for the cell. Hence, to ensure mass conservation,  condensed volatiles at the surface are computed with :
\begin{equation}
     X = \sum_{i=1}^7 X_i \frac{\delta_i}{\cos(\mu_i)}
     \label{eq:traceurs}
\end{equation}
\noindent where $\mu_i$ is defined in Table \ref{Table:Statistique}.

For surface temperature, we assume that we can average the Stefan–Boltzmann function:

\begin{equation}
    \epsilon \sigma T_{\rm{surf,grid}}^4 = \sum_{i=1}^7 \epsilon_i \sigma T_{\rm{surf,i}}^4 \delta_i
    \label{eq:Tsurf}
\end{equation}

\noindent where $\epsilon$~(unitless) is the grid emissivity, $T_{\rm{surf,grid}}$~(K) the grid surface temperature, and quantities with subscript $i$ refer to the same quantities but for the sub-grid slope.

\subsection{Post-Processing and Interpolation \label{ssec:interp}}
The proposed model allows, for each grid point, to determine the surface fields for the seven types of slopes modeled here (Figure \ref{fig:cartoon_sub-gridslopes}, Table \ref{Table:Statistique}). The question is now: for any point on the Martian surface, knowing the characteristics of the terrain (slope and azimuth), can we determine the values of the surface quantities (e.g., \CO frost, \HHO frost, surface temperature) using the GCM outputs that are discretized in space and slopes?

 Let us consider a point (noted $M$) of the surface of Mars, placed on a surface of slope angle $\theta$ and azimuth $\psi$. We  compute $\mu$ and find the index $i,i+1$ of the two GCM sub-grid slopes such that:
\begin{equation}
    \mu_i  \leq \mu < \mu_{i+1}
\end{equation}

\noindent The point $M$ is surrounded by four GCM meshes. We can interpolate the values obtained at the GCM points for the sub-grid slope  $\mu_i$ and  $\mu_{i+1}$ at point $M$. This is performed by using a bilinear interpolation in space. After that, the value for slope $\mu$  at $M$ is obtained with linear interpolation between the fields obtained after the bilinear interpolation for  $\mu_i$ and $\mu_{i+1}$. If $\mu < \mu_0$ or $\mu > \mu_7$,  an extrapolation is computed.  Thus, we have virtually access to any surface field for any kind of sloped surface on Mars using this complete climate model.  We acknowledge that 1) we assume linear interpolation on processes that are not necessarily linear (although we interpolate neighboring points); 2) we neglect the time-shift (generally lower than 1-2~hours) for slopes oriented East-West (Figure \ref{fig:insolpente}c).

\subsection{Discussions on the Modeling Assumptions \label{ssec:limitsmodel}}
\subsubsection{East-West component of the slope \label{ssec:azimtuh}}

We have demonstrated in section \ref{sec:ConstructionStatistique} that  our parameterization  is suitable to represent the diurnal (minimum, maximum, and daily mean) and seasonal variations of the (sub)surface temperatures  of any slopes through its projected slope $\mu$. However, the parameterization is not adapted to reproduce the instantaneous (sub)surface temperature for a slope with an East-West component. As shown in Figure \ref{fig:insolpente}c, d, the time shift created by the azimuth of the slope is not reproduced when simulating the thermal behavior of the slope with the parameter $\mu$. This will  impact the comparison with surface temperature measurements  (section \ref{ssec:valTsurf}:  the impact can be as high as $\pm$~10-15~K) (Figure \ref{fig:insolpente}c).  

\subsubsection{Effect of surrounding topography}
It is assumed for the computation of the reflected flux that for a given sub-grid slope, the rest of the mesh is flat. We acknowledge here that this assumption is simplifying but considering non-flat surrounding terrains would generate a much  more sophisticated form factor calculation \cite<e.g.,>{Dozier1990}, as well as a radiative coupling problem between all sub-grid surfaces \cite<e.g., section 2. of>{Schorghofer2022}. Furthermore, the model neglects  shadowing created by surrounding slopes. Hence, cold traps found in alcoves for instance \cite{Schorghofer2020} are not well represented. 

\subsubsection{Sub-grid surface thermophysical properties \label{ssec:discusshetereogenenity}}
We assume that the  thermophysical properties (albedo, emissivity, thermal inertia, see Table \ref{Table:PCMinput}) of the coarse-resolution main grid apply to the sub-grid surfaces (if they are frost-free). However, surface heterogeneities have been observed at $\sim$~km scale \cite<e.g.,>{Ahern2021}, which can lead to significant differences in the surface temperatures. 
For instance, an error of $\sim$~100~J~m$^{-2}$~K$^{-1}$~s$^{-1/2}$ on the thermal inertia can lead an error of 10~K on the surface temperature \cite<Figure 1 of >{Putzig2007}; an error of $\pm$0.1 on the albedo or $\pm 0.05$ on the emissivity would lead to an error of less than 5~K  \cite<Figure 1 of >{Putzig2007,Bandfield2008}. Nevertheless, our assumption seems reasonable as no clear constraints exist for the slope surface properties: some exhibit dust/sand \cite<e.g.,>{Tebolt2020} while others expose bedrock/high thermal inertia material \cite<e.g.,>{Edwards2009}. 

\subsubsection{Turbulent exchanges and common atmosphere between the sub-grid surfaces \label{ssec:turbdiscuss}}

We assumed in the parameterization that the turbulent exchanges are done with the predominant sub-grid surface and that all sub-grid surfaces share the same atmosphere. We acknowledge here that this scientific can be discussed. For instance,   Earth climate models have shown that turbulent mixing processes can be significantly different when sub-grid variability is considered \cite{Mahrt2000,Molod2003,deVrese2016}. However, on Mars unlike on Earth, surface-atmosphere thermal exchanges are strongly dominated by radiative fluxes (computed for each slope) rather than turbulent sensible heat flux \cite{Read2017}.

The assumption that sub-grid surfaces share a common atmosphere column can also lead to inaccuracies, particularly in near-surface thermal environments. The near-surface atmosphere energy budget is primarily controlled by the radiative process associated with the slope over the first 500~m (i.e., reflected irradiance and atmospheric absorption).  Above 500~m, the increasing thermal gradient leads to the dominance of convection for the next several   kilometers \cite{Read2017}. As a result, the first few hundred meters of the atmosphere may be colder over a poleward-facing slope than over a flat area.  It should be noted that even if convection is not dominant, it would also alter the near-surface environment, and help to homogenize the atmosphere \cite{Colaitis2013}. However, building a parameterization for this convective activity for $\sim$~km slope, i.e., in the 'grey zone' \cite<or 'Terra Incognitae'>{Wyngaard2004} for resolved/parameterized convection would require a dedicated study using Large Eddy Simulations and is out of the scope of this paper.

The infrared emission by the atmosphere comes mainly from altitudes between 2 and 10~km \cite<Figure 2 of >{Dufresne2005}.  The atmosphere at these altitudes is not influenced by small slopes \cite<less than 1~km in height difference, as observed on craters for instance, e.g.,>{Vincendon2010,Vincendon2010water} because it is mixed by winds at altitudes that do not see this small relief. For these terrains, the approximation of a shared atmosphere for the calculation of the infrared flux is therefore valid. For more significant slopes (important differences in level or length of several tens of km, e.g., canyon), this hypothesis is more questionable in addition to the non-consideration of the slope winds which might play a major role.   Furthermore, our approach assumes that the winds will effectively mix these air masses such that one can adopt a common atmospheric column for all sub-grid surfaces in the analysis. Indeed, the observed Martian slopes in our study tend to be a few kilometers wide and are separated by distances on the order of tens of kilometers.  The near-surface winds have speeds of several meters per second which can mix the atmosphere between slope regions over periods of nearly tens of minutes. With our physical time-step of 15~minutes, we assume that  this time is enough  to mix the air masses and have a common atmospheric column between the sub-grid slopes.

\subsubsection{Effect of slope winds on the sensible and latent heat \label{ssec:discussslopewind}}
Because of the low density and heat capacity of the Martian atmosphere, the sensible and latent heat terms  generally lead to negligible contributions compared to the radiative fluxes in the Martian surface energy budget \cite{Savijrvi2010, Martinez2014}. However, \citeA{Spiga2011} have demonstrated that during the night, katabatic winds can locally reinforce the sensible heat flux which becomes comparable to the thermal infrared cooling of the surface. Anabatic slope winds could also explain colder than expected surface temperatures during the afternoon \cite{Spiga2011}.

Slope winds have also an effect on the sublimation of \HHO as it is controlled by the wind velocity (Eq. \ref{eq:sublimwaterice}). As slope wind speeds can be $\sim$10 times higher than those for a flat surface, they can have great importance when modelling \HHO ice stability. Experiments using the 1D model showed that high  slope winds could reduce the ice stability by tens of sols \cite{Vincendon2010water}.

 Subgrid-scale parameterization of slope winds deserves a dedicated study and is left as a future work on how its impact on slope microclimates can be represented in Global Climate Models.

\subsubsection{Interpolation sub-grid surfaces and grix box average}

Grid-box average (section \ref{ssec:gridbox}) or post-processing \ref{ssec:interp}) are done using (bi)linear interpolation or linear average, weighted by the cover fraction. However, such methods are applied to processes that might be non-linear. These interpolations/communications between sub-grid heterogeneity and GCM grid box have been studied and reviewed in the Earth climate community \cite<e.g.,>{Giorgi1997} and future improvements of our parameterization should focus on this point.

\section{An illustration in the Olympus Mons Region \label{sec:illustration}}

\begin{figure}[!ht]
    \centering
    \includegraphics[width = 1\textwidth]{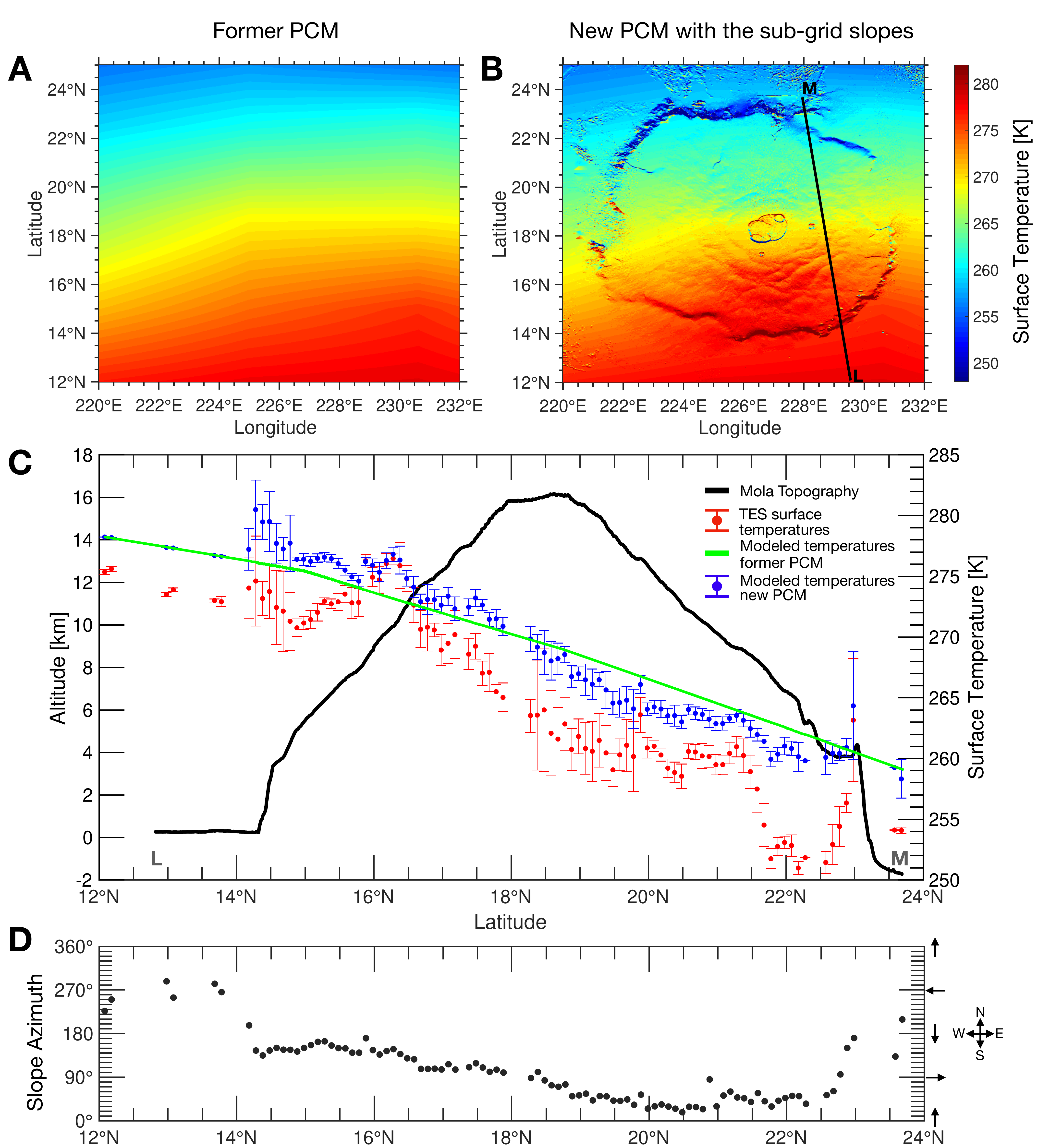}
    \caption{Surface temperatures predicted by the PCM in the Olympus Mons region at \Ls~=~270\textdegree~(\Ls~being the solar longitude), at 2~p.m for a) the PCM without the sub-grid slope parameterization b) the PCM with the sub-grid slope parameterization. Maps have a resolution of 128 pixels per degree. No MOLA background has been added to the plot. c) Comparison between the surface temperatures measured by TES (red dots) and prediction by the former PCM (green line) and new PCM (blue dots) on transect L-M (panel b). PCM predictions and TES measurements are binned: dots represent the mean value, and the error bars represent their standard deviation ($\sigma$). d) Azimuth of the slopes where TES measurements are made.}

    \label{fig:Tharsis}
\end{figure}

Here we illustrate the improvements achieved with this sub-grid slope parameterization by giving an example in the region of Olympus Mons (12\textdegree-25\textdegree N, 220-235\textdegree E). Figure \ref{fig:Tharsis}a illustrates the surface temperatures predicted by the PCM without the parameterization. In this version,  surfaces modeled by the PCM are only flat, so no temperature variation related to topography can be detected.  Figure \ref{fig:Tharsis}b illustrates the surface temperatures predicted  by the PCM with the sub-grid parameterization  in this same region, after  the post-processing described in section \ref{ssec:interp}. With this version, the obtained "thermal image" and the surface temperature distribution of the area are more realistic:  surface temperatures vary with the topography and the slopes. The reliefs of Olympus Mons  appear naturally because of the topographic gradients that create surface temperature differences. We recover expected observations: the flanks of Olympus Mons oriented toward the equator are warmer than those oriented towards the North. The same conclusions for the caldera of the volcanoes can be drawn. Similar observations are drawn when repeating this experiment in different regions (e.g., Valles Marineris, Tharsis, etc.).   This example is used for the validation in the next section.

\section{Validations of the Implementation \label{sec:Validation}}
\subsection{Validation of the Surface Temperature \label{ssec:valTsurf}}

We compare here the surface temperatures predicted by our model on slopes with observations on warm (daytime), intermediate (dawn/sunset), and cold (night/shadow) surfaces. To do so, we used surface temperatures measured by TES and the Thermal Emission Imaging System \cite<THEMIS,>{Christensen2004} infrared imager. The TES instrument  includes  a spectrometer in the 6-50~$\mu$m band and co-aligned visible (0.3 - 2.7~$\mu$m) and thermal (5.5-100~$\mu$m) bolometers  providing measurements with a $\sim$ 3-8~km resolution in a $\sim$380~km footprint. \cite{Christensen2001}. THEMIS  includes a thermal bolometer in the 7-15~$\mu$m band. and provides surface temperatures measurements  with a resolution of  $\sim100$~m within  a footprint of $\sim 32$~km. In this study, we use TES kinetic surface temperatures derived  from spectral measurements because of their low sensibility to  atmospheric effects \cite{Bandfield2008}. Furthermore, the emissivity used for TES retrievals is equal to the one of the PCM.   THEMIS measures spectral radiance emitted by the surface and converts it to brightness temperature assuming a surface emissivity of one and without applying atmospheric correction \cite{Christensen2003}. Therefore, to mitigate these effects, we only considered  surface temperatures acquired by THEMIS after 6~p.m. and before 8~p.m., or between  6~a.m. and 8~p.m. At these local times, simulations with the PCM indicate that the temperature contrast between the atmosphere and the surface should be low, so atmospheric correction, which is not done during THEMIS post-processing, should be low. Furthermore, we assume that at these times, the surface has cooled enough so that the effects of albedo variability during the day on temperatures have disappeared. We also assume that just after sunset,  the effect of thermal inertia heterogeneity on slopes is not yet significant enough to introduce a bias during the comparison. For this last point, observations at 5~p.m. would have been preferable but the effect of slope azimuth on surface temperatures would still have been marked at these hours.  Since these temperatures are calculated assuming an emissivity of 1  and the PCM assumes an emissivity of 0.95  we corrected these temperatures by a factor $0.95^{-1/4}$. Lastly, as THEMIS has a thin resolution, we have degraded the images to a resolution of 1~km, a resolution similar to the  parameterization used in this section (MOLA file with a resolution of $\sim$1~km at the equator to generate the sub-grid statistics).

\subsubsection{Daytime surface temperatures}
For the daytime comparison, we looked at the Olympus Moons region, which has  steep slopes and a surface with homogeneous properties.  We present  a comparison at \Ls~=~270\textdegree~ (Northern autumn) at 2~p.m in Figure \ref{fig:Tharsis}c.   TES measurements presented here have been made during Martian Year 26 on the transect AB (Figure \ref{fig:Tharsis}b). For each measurement, we retrieved the slope angle and azimuth associated and computed the predicted surface temperatures using the method presented in section \ref{ssec:gridbox}. We then binned the measurements and simulated temperatures  on 0.1\textdegree~latitude segments. As expected, the distribution of the surface temperatures with the parameterization is better (mean error of  -4.2$\pm 2.3$~K at \sigO) compared to the former PCM (mean error of  -4.5$\pm 3$~K at \sigO).  The model performs less well (error of $\sim$~5~K) when the slope has a strong east-west component (e.g., latitude s 16-19\textdegree N, Fig. \ref{fig:Tharsis}d). This is due to the choice of representing the slope by its meridional component, which therefore does not lead to a satisfactory simulation of the instantaneous temperature for this type of slope. Surface temperatures are overestimated by the PCM, especially on the flank of the volcano where the steepest slopes are found. This might be caused by the katabatic slope winds (not represented in the model) which should cool down  the atmosphere and decrease the surface temperatures \cite{Spiga2011}. Other surface temperature transects investigated show similar results in this region, as well as in Valles Marineis.

\subsubsection{Sunset surface temperatures}

\begin{figure}[!ht]
    \centering
    \includegraphics[width = 0.9\textwidth]{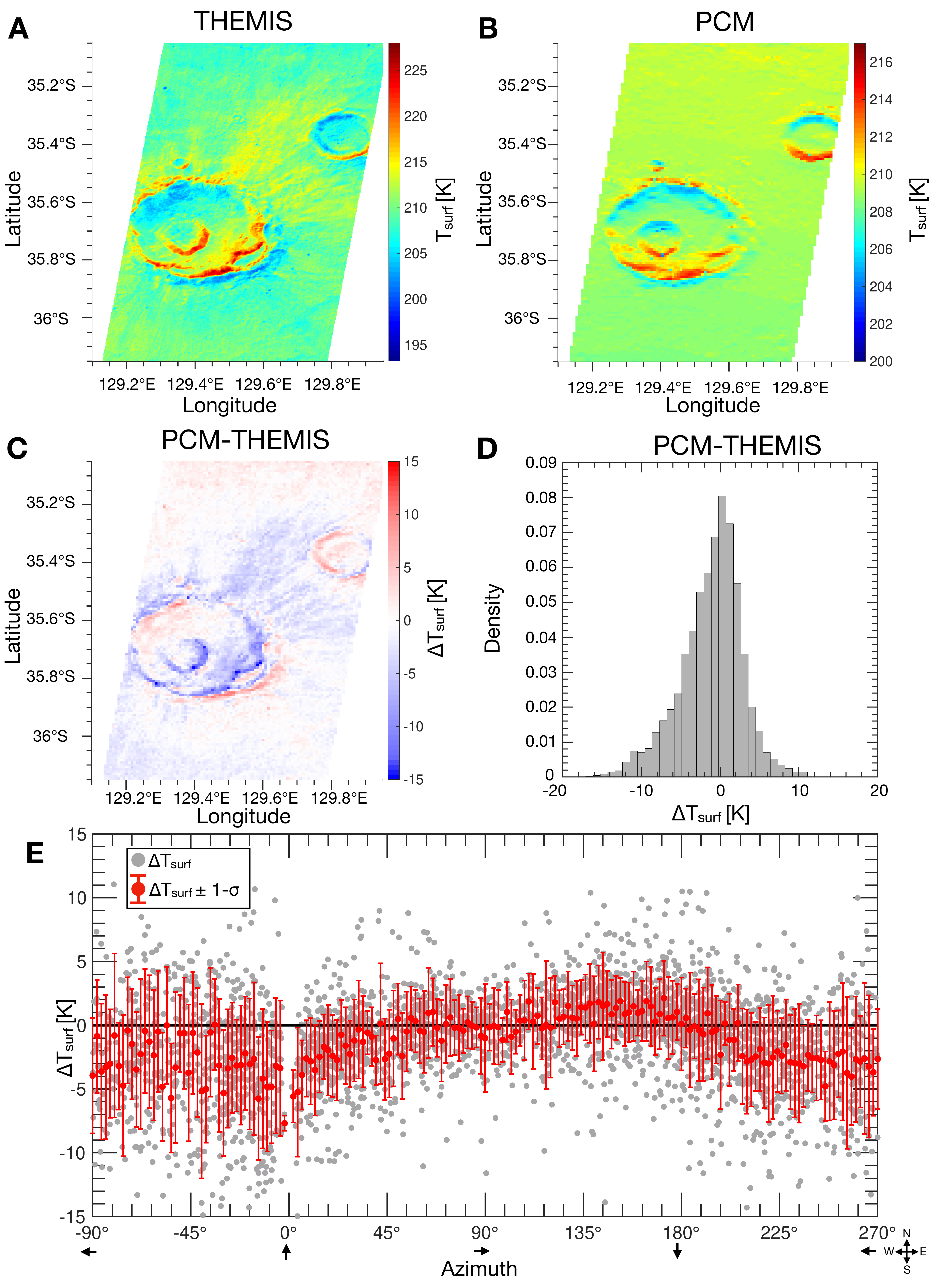}
    \caption{a) Surface Temperature obtained by THEMIS around Gasa crater on image I89710003 (35.7\textdegree S, 129.4\textdegree E) obtained at \Ls = 181\textdegree~and local~time~=~7.08~p.m.  b) Surface Temperature predicted by the PCM with the sub-grid slope parameterization. c) Difference between the PCM and THEMIS measurements d) Histogram of the difference between the PCM and THEMIS measurements on steep (slope angle $\geq$9~\textdegree) slope. e) Difference between the PCM and THEMIS measurements on slopes with a slope angle higher than 9\textdegree~ versus the slope's azimuth.  }
    \label{fig:THEMIS_TsurfI8}
\end{figure}

We then looked at temperatures acquired at sunset, which are intermediate during the warm temperatures of the day and cold ones at dawn.  An example of  validation is presented in Figure \ref{fig:THEMIS_TsurfI8}. Here, we looked at the area at Gasa crater (35.7\textdegree S, 129.4\textdegree E) during Southern spring.  In this region of interest (ROI, defined by the boundaries given in Figure \ref{fig:THEMIS_TsurfI8}a),  the majority of the surfaces  is flat, and the steep terrains are confined to the craters where significant slopes can be found ($\delta_4$~=~85\% here). THEMIS measured surface temperatures that are on average $\sim$210~K, mostly on flat surfaces (Figure \ref{fig:THEMIS_TsurfI8}a) with extreme temperatures found on East-South and West-Equatorward  facing slopes (between 200~K and 227~K respectively). 
The former PCM, without the sub-grid slope parameterization, i.e., which computes surface temperature only for flat terrains, gives a mean surface of $\sim$209~K, in good agreement with THEMIS observations but was not able to reproduce the cold and hot temperatures found on the slopes. The new parameterization helps to reproduce a realistic field where thermal contrasts created by the topography are apparent (Figure \ref{fig:THEMIS_TsurfI8}b). The difference between the PCM and THEMIS surface temperatures  (Figure \ref{fig:THEMIS_TsurfI8}c) reveals two interesting features. First, some differences exist between the PCM and the observations on flat surfaces (e.g., 35.2-35.5\textdegree S, 129.4-129.6\textdegree E). These small differences are due to sub-kilometer variability in the thermal inertia that is not caught by the PCM. Second,
the PCM seems to  underestimate the surface temperatures on equatorward-facing slopes compared to THEMIS, whereas it overestimates it for pole-facing slopes (Figure S1a). However, Figures \ref{fig:THEMIS_TsurfI8}c, e show that the most significant differences are found on East-West slopes (Figure \ref{fig:THEMIS_TsurfI8}c, e), where the PCM can be 4~K ($\pm 4$~K at \sigO) colder for a North-East-facing slope, or ($\pm 2$~K at \sigO)~K warmer for a South-West facing. On the contrary, for North-South facing slopes, the differences are limited to less than 1-2~K~($\pm 2$~K at \sigO). Again, this is due to the slope parameterization, which does not model the slope's azimuth effects (section \ref{ssec:azimtuh}). Thermal inertia effects also contribute to  the difference between the observed and simulated temperatures as the slopes presented here present a higher thermal inertia ($\geq$~1000 J~m$^{-2}$~K$^{-1}$) than the one used in the PCM (210 \tiu).  Finally, this underestimation of maximum and minimum temperatures on slopes may be linked to: 1) the common atmosphere between each sub-slope: North-facing slopes have a colder near-surface atmosphere than in reality, which could lead to an underestimation of the flow. This underestimation would be limited to 1~W~m$^{-2}$, given that the radiation comes mainly from an altitude of 2-10~km \cite{Dufresne2005}, where atmospheric temperatures are homogeneous; 2) An overestimation of a scattering component for steep slopes, as the parameterization of \citeA{Spiga2008} is indeed less efficient for such slopes.  Nevertheless, when comparing the temperatures simulated and observed on steep (slope angle $\geq$~9\textdegree) slopes (Figure \ref{fig:THEMIS_TsurfI8}d), we observed that  the modeled temperatures have an error of -1.3~K~$\pm 4$~K~at~\sigO. Other comparisons lead to other locations give similar results.

\subsubsection{Cold surface temperatures at dawn}
\begin{figure}[!ht]
    \centering
    \includegraphics[width = 0.9\textwidth]{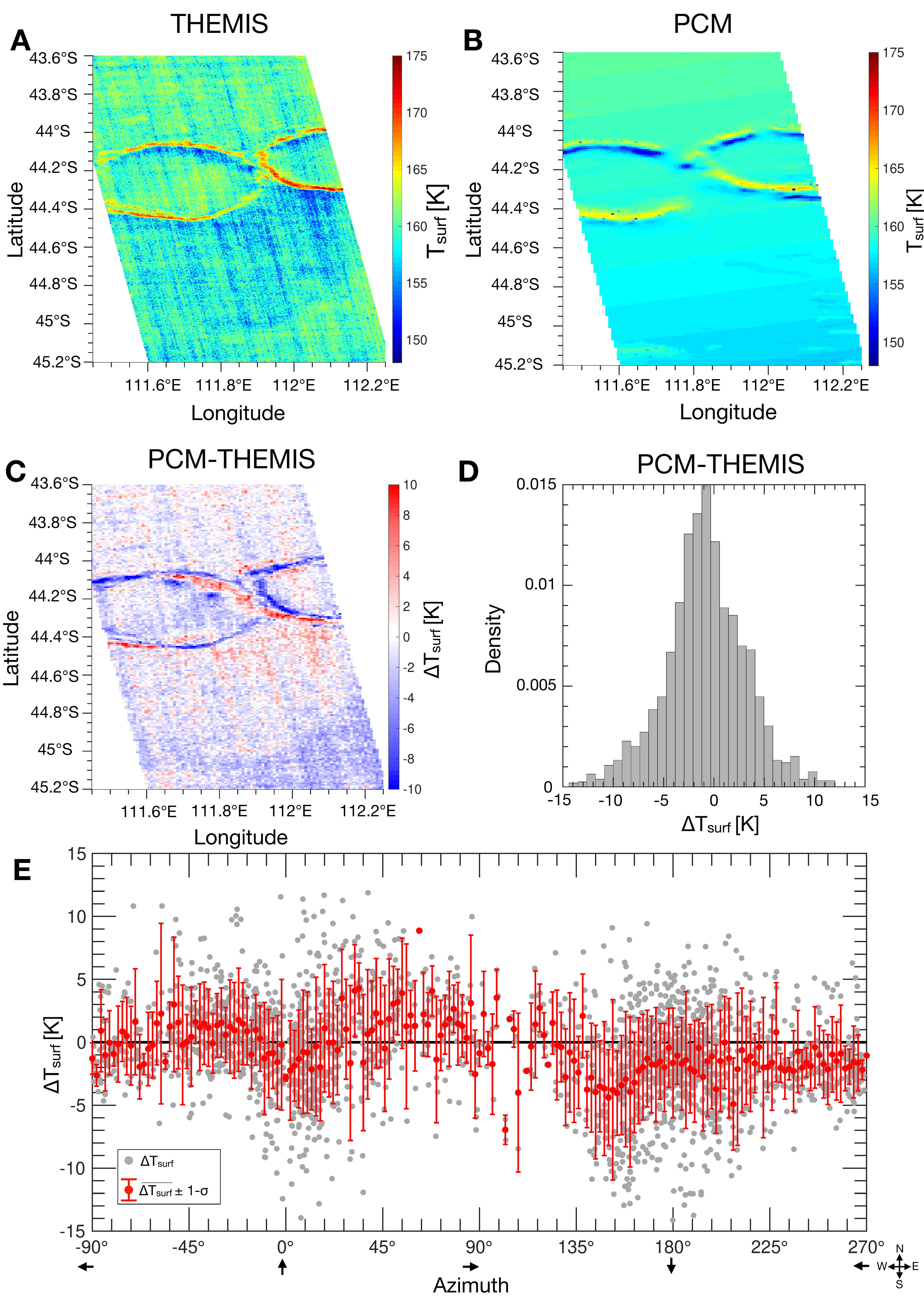}
    \caption{a) Surface Temperature obtained by THEMIS on image  I63705006  obtained at \Ls = 142.13\textdegree~and local~time~=~7.41 a.m.  b) Surface Temperature predicted by the PCM with the sub-grid slope parameterization. c) Difference between the PCM and THEMIS measurements d) Distribution of the difference between the PCM and THEMIS measurements on steep (slope angle $\geq$9~\textdegree) slope. e) Difference between the PCM and THEMIS measurements on slopes with a slope angle higher than 9\textdegree~ versus the slope's azimuth.}
    \label{fig:THEMIS_TsurfI6}
\end{figure}

We checked here our surface temperatures with observations acquired at dawn when the surface temperatures are the coldest. An illustration is given in Figure \ref{fig:THEMIS_TsurfI6}. The image was acquired at 7.41~a.m. during Southern Winter. Shadowed pole-facing slopes have a surface temperature that is within 5~K of the \CO frost point (150~K for the typical pressure at this location), suggesting the presence of  \CO ice \cite{KhullerGullies,Lange2022a}. As the emissivity of \CO ice varies with crystal size \cite[and references therein]{Piqueux2016}, we did not apply the $0.95^{-1/4}$ correction for these areas. The observation of \CO frost on pole-facing slopes at such latitudes is consistent with observations from \citeA{Vincendon2010,KhullerGullies,Lange2022a}. The same observations as those described previously are found here: sub-grid differences between the model and the observations that are created by sub-grid differences of thermal inertia, and the influence of the slope azimuth. When comparing the slope surface temperatures (Figure \ref{fig:THEMIS_TsurfI6}d, Fig S1b), we found that the model is colder than the observations (-1.1~$\pm$~5~K at \sigO). The difference varies with the slope's orientation (Figure \ref{fig:THEMIS_TsurfI6}e), with again, pole-facing slopes being warmer and equatorward facing slopes being colder by 2-3~K~($\pm$ 4~K).

\subsubsection{Generalization and conclusion}

\begin{figure}[!ht]
    \centering
    \includegraphics[width = 1.1\textwidth]{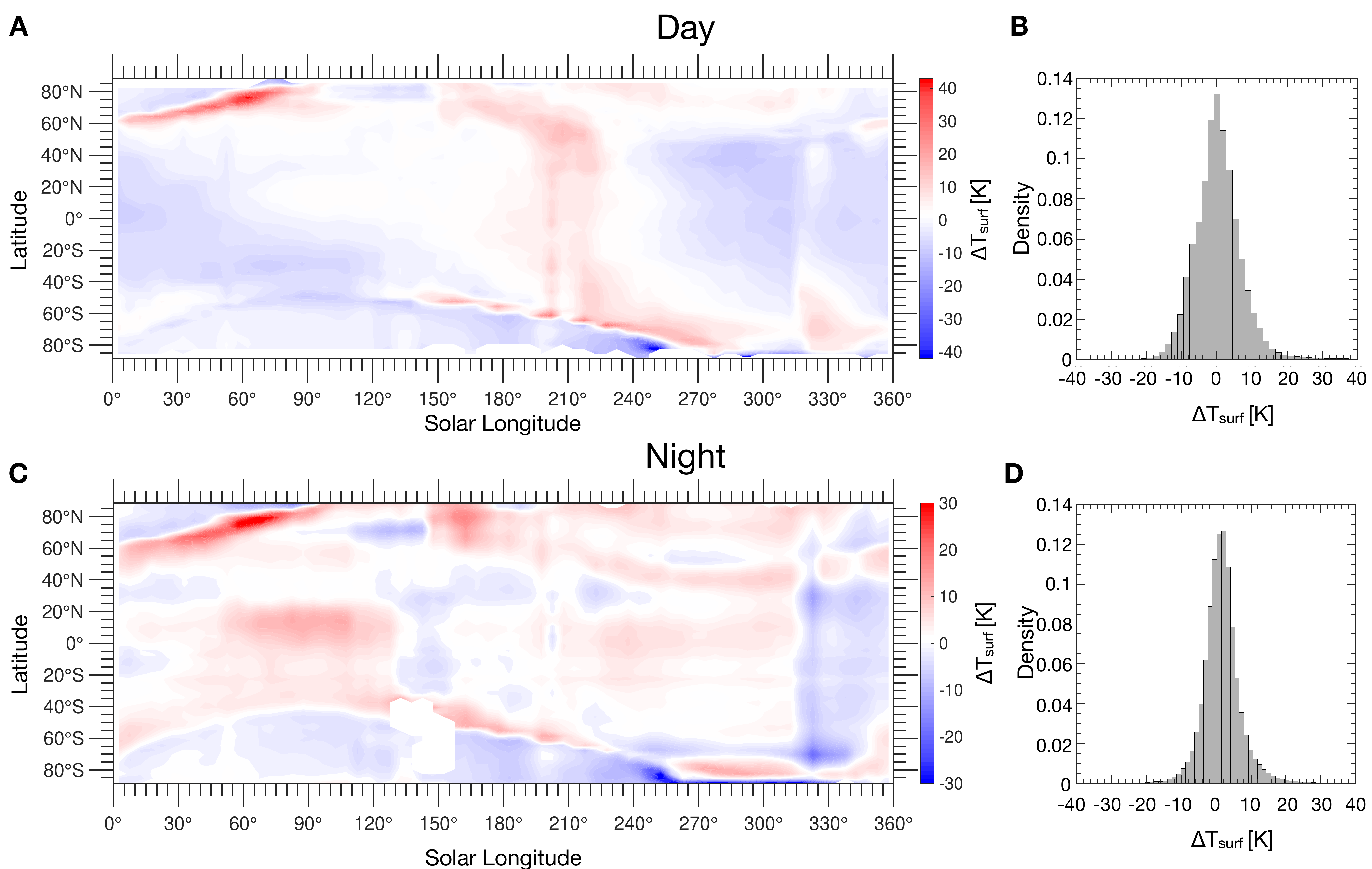}
    \caption{a) Comparison between the zonally averaged surface temperatures made by TES during MY~26 with grid-box averaged surface temperatures computed by the PCM during daytime (2~p.m.). b) Distribution of the difference between the PCM and TES. c) and d) same but for  the night (2~a.m.). The $\pm$30~K errors in the polar regions are related to the \CO caps that retreat earlier in the PCM compared to the observations. }
    \label{fig:TES_year}
\end{figure}

We extend the study presented above to other regions of Mars. Similar conclusions to those presented above can be drawn. The average surface  temperatures predicted by the PCM and what is observed by THEMIS are consistent, even though small offsets can appear.  We have quantified this difference over the year and planet by comparing the zonally average surface temperature measured by TES over MY~26 \cite{Smith2004} to the values computed by the PCM. As TES values in \citeA{Smith2004} are binned by 3\textdegree~$\times$~3\textdegree~of latitudes and longitudes, and 5\textdegree~of \Ls, we have compared these values to the surface grid-box averaged values given by the PCM. The comparison for daytime (2~p.m.) and nighttime (2~a.m.) is presented in Figure \ref{fig:TES_year}.

The PCM  generally slightly overestimates  surface temperature by $\sim 3-4$~K during summer and underestimates by $\sim 3-4$~K during winter.  On average, during the year, the PCM marginally overestimates the surface temperature by +0.5~$\pm$6~K~at~\sigO  during the day, and by 1.5~$\pm$6~K~at~\sigO during the night.  This difference was also present in the former PCM without the parameterization and does not come from our work. This  might be related to the vertical representation of dust/water in the PCM as the temperature differences (outside of polar regions) are mostly found during periods of high opacity (e.g., \Ls = 180\textdegree-210\textdegree, Figure \ref{fig:TES_year}a)). Such conclusions can be extended to slope temperatures. Previous sections have shown that these temperatures were very sensitive to the slope azimuth and its thermal inertia.  Nevertheless, we found a good agreement between our model and the observations. The agreement is better for short-scale topography (e.g., small crater, Figures \ref{fig:THEMIS_TsurfI8}, \ref{fig:THEMIS_TsurfI6}) than significant topographic gradients (e.g., Valles Marineis, volcanoes, Figure \ref{fig:Tharsis}) where slope winds can significantly impact the slope energy budget \cite{Spiga2011}.

These tests validate the overall good agreement between our model and the slope temperatures. However, we acknowledge that we cannot fully validate the surface temperatures on all slopes with our parameterization because of the difficulty of correcting the phase shift induced by the East-West components of the slopes and the under-mesh heterogeneity for the surface properties which can lead to differences up to 10-15~K between the model and the observations as presented by Figures \ref{fig:Tharsis}, \ref{fig:THEMIS_TsurfI8},  \ref{fig:THEMIS_TsurfI6} which are representatives of the error made. Again, the purpose of our parameterization is to represent the average thermal behavior of the slopes (diurnal average temperatures, minimum and maximum temperatures) and cannot fully simulate the instantaneous behavior. This is why we validate in the next section our parameterization with the observation of seasonal frost, which can theoretically be correctly represented by the parameterization.

\subsection{Modelling the Distribution of Seasonal Frost on Slopes \label{ssec:validationfrost} }
\subsubsection{Temporal Distribution}

\begin{figure}[!ht]
    \centering
    \includegraphics[width = \textwidth]{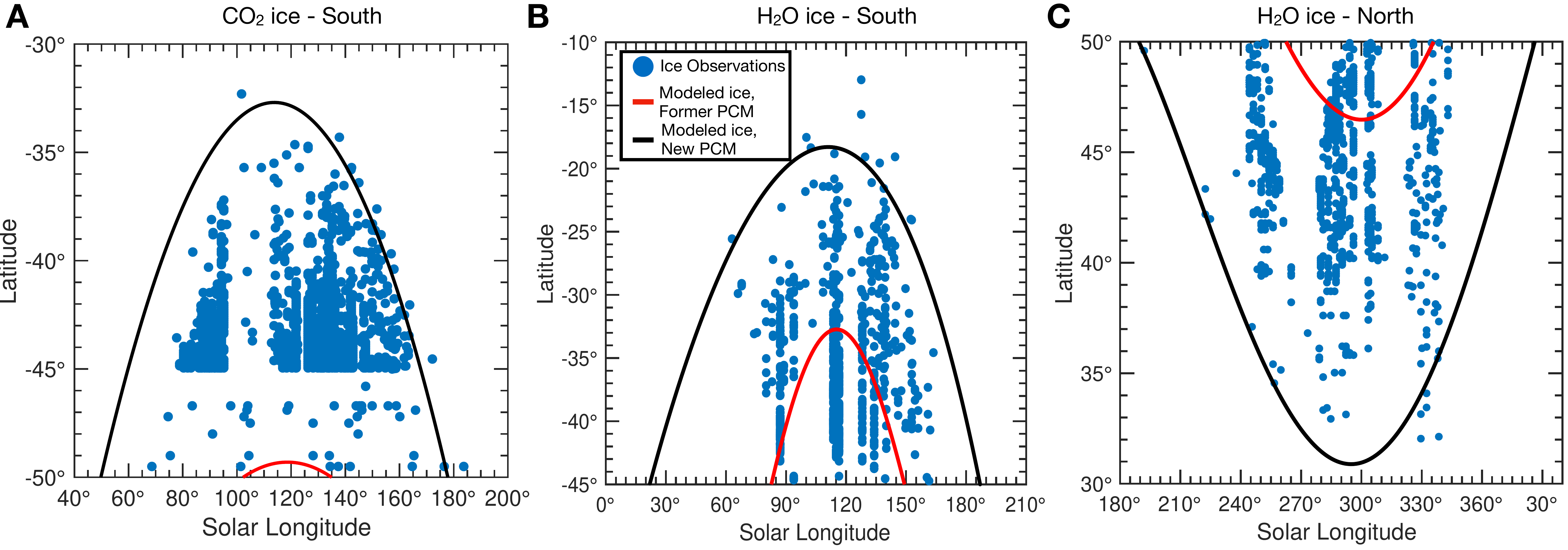}
    \caption{Latitudinal distribution of a) \CO  and b,c) \HHO frost  vs. solar longitude \Ls . Blue points correspond to the observations of frost by CRISM/OMEGA \cite{Vincendon2010, Vincendon2010water, Vincendon2015}. The red curve is the prediction of frost stability by the former PCM without the sub-grid slope parameterization, and the dark curve is the prediction of frost stability on a 30\textdegree~pole-facing slope.  Maximum extent of \CO ice predicted  by the PCM in the East of Hellas (100-150\textdegree E). \CO ice is predicted if the  of \CO frost thickness exceeds $ 200 \mu$m at \Ls~$\geq$~120\textdegree, 1000~$\mu$m before. \HHO ice is predicted at a given latitude by the PCM if the \HHO frost thickness exceeds $ 5 \mu$m.  \HHO frost shown in b) is located in the band 0\textdegree–60\textdegree E, and in the band  280\textdegree E–340\textdegree E for c). Similar conclusions can be drawn at other longitudes.  The PCM outputs are retrieved at 2 p.m., i.e., the local time where most CRISM/OMEGA data were acquired }
    \label{fig:DistributionLsFrost}
\end{figure}

We now compare the distribution of seasonal \CO and \HHO frosts on slopes with the predictions of the PCM. For this validation, we used the frost detections made by  OMEGA (Observatoire pour la Minéralogie, l'Eau, les Glaces et l'Activité) and CRISM (Compact Reconnaissance Imaging Spectrometer for Mars) instruments as reported in \citeA{Vincendon2010,Vincendon2010water,Vincendon2015}. These detections were made using  near-infrared spectroscopy in the afternoon, and correspond to seasonal frost. Hence, diurnal frost forming during the night and sublimating during the early morning is not considered \cite{Piqueux2016,KhullerGullies,Lange2022a}. These observations are compared with the prediction of \CO and \HHO ice stability on a 30\textdegree~pole-facing slope retrieved from the PCM simulations at the same local time, i.e., 2~p.m. Following \citeA{Vincendon2010}, we assume that \CO ice is stable and should be detected by CRISM/OMEGA at a given latitude if the  of \CO ice thickness predicted on the slope by the PCM  exceeds $ 200~\mu$m at \Ls~$\geq$~120\textdegree, 1000~$\mu$m before. The threshold used is higher for \Ls~$\leq$~120\textdegree~because of the low signal-to-noise ratio on the slope at this time of the Martian year. Imposing a lower threshold (e.g., 100~$\mu$m) would lead to an earlier condensation by $\sim$5 degrees of \Ls, and an equatorward extent of the distribution  by 3\textdegree~of latitude. For  \HHO ice, the thickness threshold is set to $ 5~\mu$m \cite{Vincendon2010water}.  The comparison is presented in Figure \ref{fig:DistributionLsFrost}.

\paragraph*{CO$_2$ ice:}

The PCM with the sub-grid slope parameterization predicts \CO ice on  30\textdegree~pole-facing slopes in the southern hemisphere up to 33\textdegree S while the former PCM could predict \CO ice on flat surfaces as low $\sim 48$\textdegree S. This new latitudinal extent is consistent with observations by CRISM and OMEGA which detected \CO frost on south-facing slopes up to 32.3\textdegree S.  We observed an overall very good agreement between our modelling predictions and actual CO$_2$ detections.

The timing of condensation/sublimation is consistent with the observations, even though both start  slightly too early~(by 5-10\textdegree). \citeA{Vincendon2010} showed that the beginning of condensation is sensitive to the amount of dust in the atmosphere, the presence of ice in the subsurface, and the thermal inertia of the surface. Dust is set to the observations of \citeA{MONTABONE2015} and can not be tuned to match the observations. The subsurface ice is predicted to be stable at a depth of $\sim 1$~m in the 35\textdegree-45\textdegree S-band (see section \ref{ssec:groundice}). At such depth, subsurface ice has almost no impact on the surface energy budget and \CO ice distribution.  Hence, the difference between our model and the \CO ice observations suggests shallower ice that is not predicted by our ice stability model. One can also delay the beginning of the modeled \CO condensation/sublimation by modifying slope surface properties. However, no clear constraints exist for these properties as some slopes exhibit low thermal inertia \cite{Tebolt2020}, favoring the condensation of \CO; while some slopes reveal  high-thermal inertia bedrock exposures \cite{Edwards2009}  which delay the formation of \CO frost. Finally, one should note that the difference in the timing of condensation is observed at \Ls~$\leq 120$\textdegree, a time where fewer ice observations are available because of the low illumination in this period, lowering the signal-to-noise ratio, and potentially preventing some detections.

Here, our model matches well the observations without relying on subsurface ice as in \citeA{Vincendon2010}. This may be due to the fact that our model considers  the surrounding plains that warm the atmosphere, increasing the downwelling infrared flux and therefore the surface temperature.    Such effects, which were not considered in \cite{Vincendon2010}, and their consequences are discussed in section \ref{ssec:discuss1} and will be more detailed in a companion paper \cite{Lange2023ice}.

\paragraph*{\HHO ice:} 
Similar conclusions can be drawn when looking at the predicted latitude-season stability diagram for \HHO ice (Figure \ref{fig:DistributionLsFrost}b. and c.) 
The latitudinal extent of \HHO ice is well predicted (as low as 30\textdegree N and 18 \textdegree S vs. 48 \textdegree N and 33 \textdegree S for the former PCM), even if few deposits at low latitudes (e.g., the one close in Valles Marineris at latitude 12.7\textdegree S) are not predicted by our model. In our experiment, the broadband albedo of \HHO ice is set to 0.33 as it enables a good fit of the seasonal water cycle in the PCM \cite{Naar2021}. Similarly to \citeA{Vincendon2010water}, increasing the albedo leads to a delay by a few degrees of \Ls~in the sublimation of the deposits. The sublimation of these deposits is also sensitive to wind speed. Here, this speed might be underestimated because the model does not compute slope winds although this should not significantly impact our results \cite{Vincendon2010water}. The beginning of the condensation of \HHO is also too early by $\sim$10\textdegree~of \Ls~in the PCM compared to the observations, as for the \CO ice. This might be due to an observational bias in the autumn high latitude water ice detections at these longitudes \cite{Vincendon2010water}, and a bias of our model that might underestimate surface temperature at this season as reported previously. One should note that our results are quite similar to those obtained by \citeA{Vincendon2010water}.

\subsubsection{Spatial Distribution}
\begin{figure}[!ht]
    \centering
    \includegraphics[width = 1\textwidth]{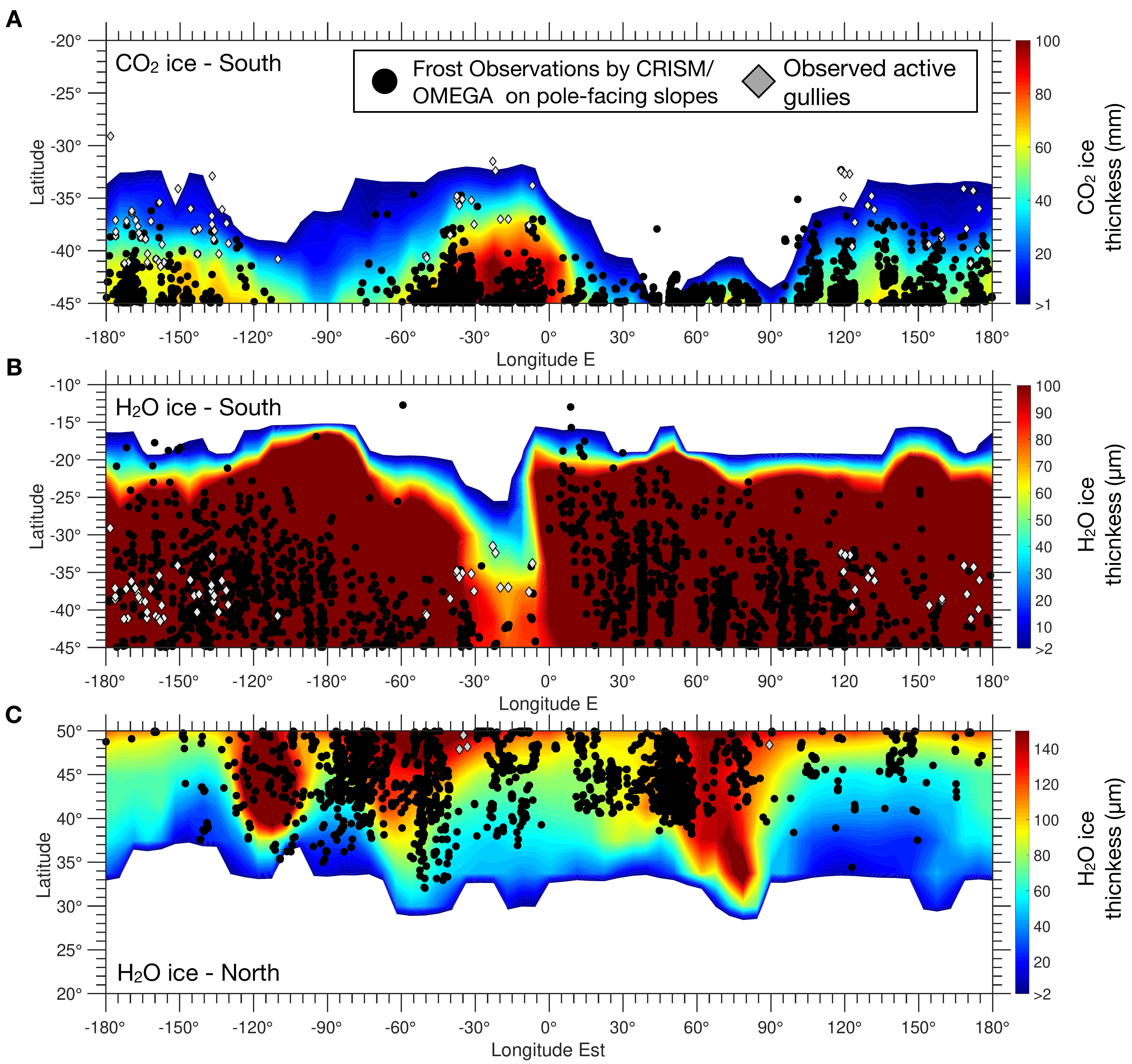}
    \caption{Spatial distribution of \CO and \HHO on a 30\textdegree pole-facing slopes predicted by our model and ice detection made by CRISM and OMEGA (black dots). a) \CO ice in the Southern Hemisphere;  b) \HHO  ice in the Southern Hemisphere; c) \HHO  ice in the Northern Hemisphere. The thicknesses presented here are the maximum frost thickness across the entire Martian year. Observed active gullies reported by \citeA{Dundas2022} are represented by gray diamonds. 
    The PCM outputs are retrieved at 2 p.m., i.e., the local time when most CRISM/OMEGA data were acquired.}
    \label{fig:map_ice_latlon}
\end{figure}

We also validate our parameterization by comparing the spatial distribution of \CO and \HHO ice observed by CRISM/OMEGA with our model. Results are presented in Figure \ref{fig:map_ice_latlon}. 

\paragraph*{\CO ice:} Similarly to \citeA{Vincendon2010}, the latitudinal stability limit of \CO ice on any slope (Figure \ref{fig:map_ice_latlon}a) varies with longitude, with a mean value of ~33\textdegree. Below 33\textdegree, ice might be present but as thin deposits ($<$~100 $\mu$m). \CO ice between -120\textdegree - -60\textdegree E, i.e., Thaumasia regions, is predicted to be thinner compared to the rest of the longitudes because of the higher altitude in this place, and thus the lower temperature of \CO condensation. Our statistics of sub-grid slopes also suggest that the lack of \CO ice detections in this region might be linked to the few numbers of steep ($>$~30\textdegree) south-facing slopes in this area. Interestingly, we found that  \CO ice detections by CRISM/OMEGA correlate with thick \CO ice  predicted by the model, while \citeA{Vincendon2010} proposed that deposits of few hundred of microns should be detected. The overall agreement validates our model for \CO deposits. Few points are not predicted (e.g., the detection at 32.3\textdegree S on the East of Hellas) but this might be linked  to favorable local conditions or the resolution of our model (3.75\textdegree~of latitude) which might be not precise enough to differentiate what happens within a very few degrees of latitude. One can note that the model does not predict ice in Hellas Basin whereas it has been observed on flat surfaces (Piqueux et al., 2015) and on the rare steep slopes present at this place (Vincendon et al. 2010a). This is because of a systematic error in the Mars PCM (also present without the parameterization of slopes) which is under investigation.

\paragraph*{\HHO ice:} Similar conclusions can be drawn when looking at the spatial distribution of water ice.
In the South (Figure \ref{fig:map_ice_latlon}b), our model reproduces well the observations, with notably the dry area in the band -40\textdegree E - 0\textdegree E where thinner frost (or no frost) is expected, as confirmed by the low amount of detections in this area. This is a consequence of the  western jet  on the eastern side of the Tharsis bulge \cite{Joshi1994,Joshi1995} leading to a dryer atmosphere (less water vapor available to form H$2$O frost) \cite{Vincendon2010water}. Again, the model predicts ice in the Thaumasia region which is not observed. The same explanations as those mentioned above for \CO ice, similar to those given by \citeA{Vincendon2010water}, can be given. Locally, the latitudinal stability is higher than what is predicted by \citeA{Vincendon2010water} (e.g., -180\textdegree E~-~-150\textdegree E) and explains some low latitude ice detections. In the North, the model predicts well the areas where numerous detections are made, as well as dry areas (e.g., -180\textdegree E~-~-150\textdegree E) where fewer ice deposits are observed by CRISM/OMEGA. 

\section{Discussions and Perspectives \label{sec:Perspectives}}
\subsection{Relationship between the atmosphere 
and the slope microclimates \label{ssec:discuss1}}

One may first wonder if the slope microclimates impact the global climate. It is assumed here that the atmosphere sees an average of the sub-grid fields, weighted by the cover fraction. In section \ref{sec:ConstructionStatistique}, we showed that the predominant sub-surface is the flat one (non-flat sub-grid surfaces represent 5\% of the overall area). A significant impact of the slope microclimates on the global climate was thus not expected. We checked this
assumption by comparing PCM runs with and without this parameterization and found little differences. For instance, the relative difference of the total mass of \CO ice that condenses during the year between simulation with and without the sub-grid slope parameterization is lower than 1\%. Similar conclusions can be drawn for the amount of \HHO ice, and the total amount of \HHO in the atmosphere. This is because 1) steep slopes, which accumulate  a larger amount of frost compared to flat surfaces,  are sparse and of limited length  \cite<tens or hundreds of meters,>{Aharonson2006}; 2) regionally, some areas with a lot of pole-facing slopes have an excess of ice when considering the parameterization, but this is balanced by areas with a lot of equatorward-facing slopes, which have lower amounts of ice.

 To date, slope microclimates have mostly been explored with 1D models \cite<e.g.,>{Aharonson2006,Vincendon2010} which neglect contribution from the large-scale atmosphere dynamic. Therefore, one may wonder about the impact of the global climate on slope microclimates. In the 1D PCM used in \citeA{Vincendon2010,Vincendon2010water}, the atmosphere is in radiative equilibrium with the surface studied. For a pole-facing slope, the atmosphere above it is therefore cold. However,  in our 3D model, we assume that the atmosphere is horizontally mixed at the scale of the slopes. The atmosphere sees an average of sloped and flat sub-grid surfaces. As steep sloped terrains represent actually a  small percentage of the overall surface of Mars (see section 2), the atmosphere is mostly in equilibrium with the warmer flat sub-grid surface.  Hence, in the 3D model, the air close to cold sloped surfaces is warmer than in the 1D model as it is warmed by the nearby warm plains. This legitimates the computation of other models \cite<e.g.,>{Mellon2001,Aharonson2006,Schorghofer2006,Kieffer2013} which used the same atmospheric infrared radiation for flat surfaces and steep slopes. Also, during winter, warm air is brought above pole-facing slopes because of the subsidence from the Hadley cell at mid and subtropical latitudes, which leads to adiabatic heating of the atmosphere  \cite{Barnes2017}. This warmer atmosphere increases the downwelling infrared flux and thus modifies the amount \CO/\HHO ice condensing on the surface. Note that this effect is mostly pronounced for areas with a strong thermal contrast between the slopes and the flat surfaces and is thus most significant for low-latitude slopes ($\leq$~$\pm$45$^\circ$N). These two effects increase the slope energy budget by nearly 10~W~m$^{-2}$,  with significant consequences for CO$_2$ frost extent. A more accurate description and demonstration of these two effects and their consequences are discussed in a companion paper \cite{Lange2023ice}. As discussed in sections \ref{ssec:turbdiscuss} and \ref{ssec:discussslopewind}, this assumption of a "shared" atmosphere, mostly in equilibrium with the flat surfaces, might lead to a small overestimation of the downwelling infrared flux, as cold air can, for instance, be confined within a crater by katabatic slope winds during the night and early morning. This might explain why our model has a too early starting of the \CO sublimation (Figure \ref{fig:DistributionLsFrost}) compared to the observations, but our choice is reasonable to study small/medium topographic relief.

\subsection{Reappraisal of the frost thicknesses and consequences for the activity of gullies \label{ssec:gullies}} 

Estimating the thickness of seasonal frosts (H$_2$O or CO$_2$ ice) on slopes is an important constraint for models of gully activity associated with sublimation processes \cite{Dundas2019gullies,Diniega2021}. No measurements of these thicknesses have been made and they have only been estimated by 1D models to date. Such studies only estimate frost thickness for specific locations and not at a global scale \cite<e.g., >{Schorghofer2006,Vincendon2015,KhullerGullies}. We provide in Figure \ref{fig:map_ice_latlon} a complete mapping (220~$\times$~330~km of resolution) of the expected maximum CO$_2$ frost thickness across the year for 30\textdegree~pole-facing slopes. Compared to most of the 1D models used in the literature, our model with the sub-grid slope parameterization takes into account the seasonal CO$_2$ cycle that has been validated against observations, CO$_2$ depletion at polar latitudes, and the effects of large-scale meteorology. The seasonal frost thickness logically decreases with latitude, from a few mm at 33\textdegree S ($\pm$3\textdegree~of latitude due to the model resolution) to a few tens of cm at 45\textdegree S. Equatorward of 33°S, the model predicts $\sim$0.1 millimeters frost forming during the night which disappears during the morning, as observed by \citeA{KhullerGullies,Lange2022a}. Our model predicts larger ice thicknesses for mid and high latitudes (tens of cm). Interestingly, we find that in the southern hemisphere, the amount of CO$_2$ that condenses on a 30\textdegree~pole-facing slope  has a similar frost thickness to the amount that condenses on flat terrain at about ten degrees of latitude lower. For example, at 41\textdegree S, $\sim$5~cm of frost can form on a 30\textdegree~pole-facing slope, which is the same amount as that observed at the edge of the seasonal cap at $\sim$48\textdegree S (Litvak 2004). This is not observed in the Northern Hemisphere because of the shorter winter for the current eccentricity of Mars. However, as mentioned previously, this phenomenon will not impact the global mass of CO$_2$ that will condense during the winter because these very steep slopes are rare on Mars and of limited length  \cite<tens or hundreds of meters,>{Aharonson2006}.  The longitudinal variations of frost thickness reflect the variations in surface properties (albedo, thermal inertia), and altitude. For water, we find a distribution similar to that of \citeA{Vincendon2010water} who used a 1D model coupled with precipitation maps obtained in 3D. The frost thicknesses obtained are larger in our model by ten's/hundreds of micrometers because the water cycle of the 3D model has been improved \cite{Naar2021} and is now wetter compared to the version used by \citeA{Vincendon2010water}.

These new frost maps, compared to the map of active gullies, give us important information for modeling gully activity on Mars. Indeed, near-infrared observations suggest that active gullies are highly correlated to the presence of CO$_2$ frost \cite{Vincendon2015}. By constructing high-resolution surface temperature maps with the method described in section 3.3 (Figure 10), we can see that the location of most of the active gullies indeed coincides  with slopes where CO$_2$ condensation occurs. More specifically, we found that the model predicts frost on 30\textdegree~slopes on  91\% of the active gully sites and 86\% of all gully sites in the South. In the North, this percentage is reduced to 56\% for active sites and only 9\% for all gullies. Our derived values are upper limits as the amount of ice is predicted for  steep ($\geq$30\textdegree) pole-facing slopes for which the frost is most stable. Using a lower slope angle would hence reduce the statistics presented there. Nevertheless, our numbers are consistent with the same statistics derived in \citeA{KhullerGullies}.  Hence, both observations and modeling suggest that the activity of gullies is mostly related to processes involving CO$_2$ ice. However, the model does not predict a significant amount of CO$_2$ ice (a few cm thick, Figure \ref{fig:map_ice_latlon}a), thus disfavoring models that require a significant amount of ice \cite<e.g., >[although this model is for linear gullies which are mostly found above 45° latitude, where thicker ice is expected]{Diniega2013}, and promoting processes that do not need significant amounts of ice (less than a cm) to enable fluidization \cite{deHaas2019}. Note that all active gullies closest to the equator are not necessarily correlated with the presence of CO$_2$ according to the model (although the  model might not predict CO$_2$ ice in these locations because it neglects 3D effects like shadowing for instance) and observations, but are correlated with the presence of water ice frost. Hence, the possible contribution of water ice-related mechanisms can not be ruled out. This preliminary analysis will be extended to present and past orbital configurations in future dedicated studies.

\begin{figure}[!ht]
    \centering
    \includegraphics[width = 1\textwidth]{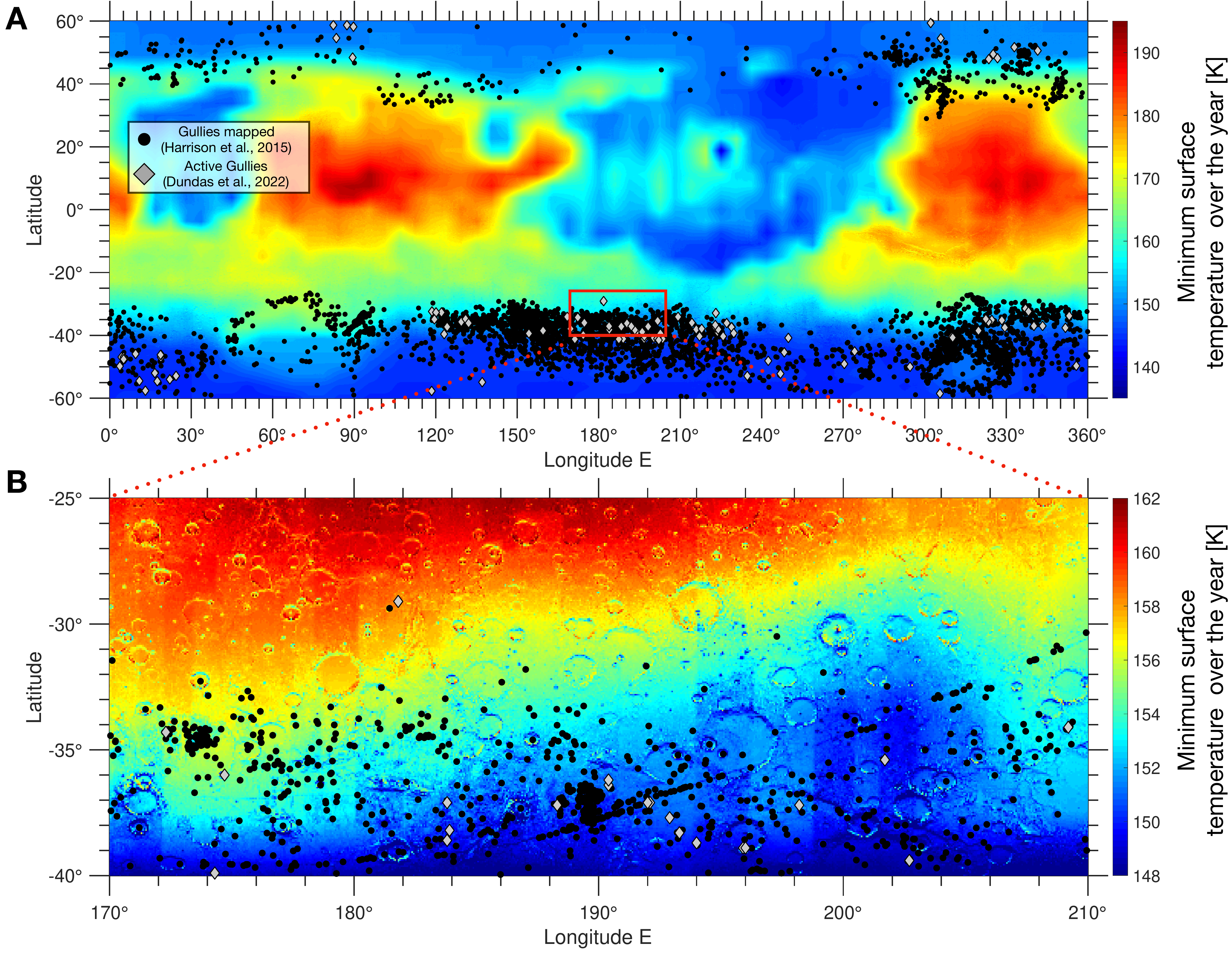}
    \caption{Minimum surface temperature computed by the PCM with the sub-grid slope parameterization between 60\textdegree S and 60\textdegree N. Gullies site mapped by \citeA{Harrison2015} are represented by the black dots, and active gullies site reported in \citeA{Dundas2022} by the grey diamonds. No MOLA background has been added to the plot. Panel B) is a close-up of panel A) between 45\textdegree S - 20\textdegree S, 170\textdegree E - 210 \textdegree E.
    }
    \label{fig:gullies}
\end{figure}

\section{Conclusions \label{sec:Conclusions}}

The objectives of this paper are to propose a parameterization to simulate slope microclimates in coarse grid GCM, implement it in the Mars Planetary Climate Model, and discuss the impact of these microclimates on the global climate of Mars and on the activity of some surface processes. The main conclusions of this investigation are: 

\begin{enumerate}
\item We can simulate the slope microclimates by representing each Mars GCM cell as a set of flat and sloped sub-grid surfaces on which we compute the surface and subsurface temperatures, as well as \CO and \HHO ice deposits (Figure \ref{fig:cartoon_sub-gridslopes}).

\item To correctly represent the properties of the sub-grid sloped surfaces, we have  demonstrated that  any given slope with a slope angle $\theta$ and azimuth $\psi$ can be on average thermally represented (same surface temperature ranges and averages) by a slope $\mu = \theta\cos(\psi)$ that is either North-facing if $\mu~>~0$, or South-facing if $\mu~<~0$ (Figure \ref{fig:insolpente}); 

\item Using seven values for $\mu$ is enough to represent accurately the sub-grid slopes (Figure \ref{fig:distribution_mu}). For the quasi-totality of the Mars PCM cells, the predominant sub-grid surface is the flat one (Figure \ref{fig:mapslope});

    \item This parameterization improves the prediction of observed  surface temperatures created by sub-grid  heterogeneities compared to the former version of the PCM (Figure \ref{fig:Tharsis}, \ref{fig:THEMIS_TsurfI8}, \ref{fig:THEMIS_TsurfI6}, \ref{fig:TES_year}).
    
    \item The spatial and temporal distribution of \CO and \HHO frost on pole-facing slopes obtained from our model agree well with measurements made by CRISM and OMEGA (Figures \ref{fig:DistributionLsFrost}, \ref{fig:map_ice_latlon}).

    \item

   While steep slopes can accumulate a significant amount of CO$_2$ and H$_2$O ice (Figure \ref{fig:map_ice_latlon}), slope microclimates do not impact the seasonal water and  CO$_2$  cycle as steep slopes are sparse on Mars. The extra amount of frost that condenses because of the sub-grid parameterization is less than 1\% of the total mass of the ice on the planet;

    \item Warm plains around steep pole-facing slopes heat the atmosphere, increasing the incident infrared flux, preventing the formation of CO$_2$ ice at low latitudes (section \ref{ssec:discuss1});

    \item We provide for the first time a  global map of CO$_2$  frost thicknesses and a revised map of H$_2$O  frost thicknesses on steep ($ \geq$~30\textdegree) slopes, which presented larger deposits than previously reported (Figure \ref{fig:map_ice_latlon}); 

    \item Our models, complemented by specific observations at near-infrared \cite{Vincendon2015} suggests that the activity of gullies is mostly related to processes involving CO$_2$ ice, but rules out processes that require a significant amount of ice (section \ref{ssec:gullies});

\end{enumerate}
This new model opens the way to novel modelling studies for surface-atmosphere interactions in the present but also past climates, allowing a better understanding of the geological record observed today. Future improvements of the model are underway, including the addition of sub-grid slope winds. Finally, our  parametrization proposed here on Mars is not specific of this planet but can be adapted to bodies whose energy balance is dominated by radiative exchanges (for instance, Pluto, Triton, etc.)

\section*{Open Research}
 THEMIS, TES infrared images, and MOLA data  can be retrieved from the Planetary Data System \cite{PDSThemisPBT,PDSTES,Smith1999} at \url{ https://pds-atmospheres.nmsu.edu}.  Frost detections are from \citeA{Vincendon2010,Vincendon2010water,Vincendon2015}  Gullies site reported in the last Figure can be obtained from \cite{Harrison2015,Dundas2019gullies,Dundas2022}.  Data files for figures used in this analysis are available in a public repository, see \citeA{LANGEJGRdataset}. The Mars PCM  used in this work can be downloaded with documentation from the SVN repository at \url{https://svn.lmd.jussieu.fr/Planeto/trunk/LMDZ.MARS/.} More information and documentation are available at http://www-planets.lmd.jussieu.fr.

\acknowledgments
This project has received funding from the European Research Council (ERC) under the European Union’s Horizon 2020 research and innovation program (grant agreement No 835275, project "Mars Through Time").  Mars PCM simulations were done thanks to the High-Performance Computing (HPC) resources of Centre Informatique National de l’Enseignement Supérieur (CINES) under the allocation n\textdegree A0100110391 made by Grand Equipement National de Calcul Intensif (GENCI). LL thanks S. Conway and A.Noblet for precious help with the gullies dataset. The authors thank G.Martinez, C.Dundas, and A.Khuller for their high-quality reviews.


%
\bibliography{agusample.bib}
\end{document}